%% file: draft_BESIII_Ds_lnu_DstDst_submit.tex
\documentclass[aps,prd,onecolumn,showpacs,amsmath,amssymb]{revtex4-1}
\usepackage{amsmath}
\usepackage{makecell}
\usepackage{graphicx}
\usepackage{sub figure}
\usepackage{epstopdf}
\usepackage{color}
\usepackage{multirow}
\usepackage{setspace}
\usepackage{overpic}
\usepackage{amssymb}
\usepackage[driverfallback=dvipdfmx, bookmarksnumbered, pdfstartview=FitH,colorlinks,urlcolor=blue, citecolor=blue,linkcolor=blue] {hyperref}
\usepackage{lineno}
\usepackage{bm}
\usepackage{rotating}
\usepackage{array}
\usepackage{threeparttable}
\usepackage[utf8]{inputenc}

\newcommand{\muv}{D_{s}^{+}\to \mu^{+}\nu_{\mu}}
\newcommand{\tauv}{D_{s}^{+}\to \tau^{+}\nu_{\tau}}
\newcommand{\tmuvv}{\tau^+ \to \mu^+\nu_{\mu}\bar\nu_{\tau}}
\newcommand{\tevv}{\tau^+ \to e^+\nu_{e}\bar\nu_{\tau}}
\newcommand{\tpiv}{\tau^+ \to \pi^+\bar\nu_\tau}
\newcommand{\trhov}{\tau^+ \to \rho^+\bar\nu_\tau}
\newcommand{\mm}{M_{\rm miss}^2}

\newcommand{\dstpdstm}{D_{s}^{*\pm}D_{s}^{*\mp}}

\newcommand{\Dstotaunue}{D^+_s\to \tau^+_e\nu_\tau}
\newcommand{\Dstotaunum}{D^+_s\to \tau^+_\mu\nu_\tau}
\newcommand{\Dstotaunup}{D^+_s\to \tau^+_\pi\nu_\tau}
\newcommand{\Dstotaunur}{D^+_s\to \tau^+_\rho\nu_\tau}

\newcommand{\muvcsresult}{0.986\pm0.023_{\rm stat}\pm0.014_{\rm syst}\pm0.003_{\rm input}}
\newcommand{\mufdsxvcsresult}{246.5\pm5.9_{\rm stat}\pm3.6_{\rm syst}\pm0.5_{\rm input}}
\newcommand{\mufdsresult}{253.2\pm6.0_{\rm stat}\pm3.7_{\rm syst}\pm0.6_{\rm input}}
\newcommand{\tauvcsresult}{1.011\pm0.014_{\rm stat}\pm0.018_{\rm syst}\pm0.003_{\rm input}}
\newcommand{\taufdsxvcsresult}{252.7\pm3.6_{\rm stat}\pm4.5_{\rm syst}\pm0.6_{\rm input}}
\newcommand{\taufdsresult}{259.6\pm3.7_{\rm stat}\pm4.6_{\rm syst}\pm0.6_{\rm input}}
\newcommand{\bftauv}{5.60\pm0.16_{\rm stat}\pm0.20_{\rm syst}}
\newcommand{\bfmuv}{0.547\pm0.026_{\rm stat}\pm0.016_{\rm syst}}
\newcommand{\Rbftauv}{5.39\pm0.14_{\rm stat}\pm0.20_{\rm syst}}
\newcommand{\Rbfmuv}{0.553\pm0.014_{\rm stat}\pm0.021_{\rm syst}}
\newcommand{\Ctauvcsresult}{0.989\pm0.006_{\rm stat}\pm0.007_{\rm syst}\pm0.003_{\rm input}}
\newcommand{\Ctaufdsresult}{254.8\pm1.6_{\rm stat}\pm1.8_{\rm syst}\pm0.6_{\rm input}}

\newcommand{\Cmuvcsresult}{0.972\pm0.009_{\rm stat}\pm0.005_{\rm syst}\pm0.003_{\rm input}}
\newcommand{\Cmufdsresult}{249.4\pm2.3_{\rm stat}\pm1.2_{\rm syst}\pm0.5_{\rm input}}

\newcommand{\Ctotfdsresult}{252.1\pm1.3_{\rm stat}\pm1.7_{\rm syst}\pm0.5_{\rm input}}
\newcommand{\Ctotvcsresult}{0.982\pm0.005_{\rm stat}\pm0.007_{\rm syst}\pm0.003_{\rm input}}

\begin{document}

\title{\bf \boldmath
Measurement of the branching fraction of $D^+_s\to \ell^+\nu_\ell$ via $e^+e^-\to D^{*+}_{s} D^{*-}_{s}$}

\input{authorlist_2024-01-03}

\begin{abstract}
Based on  $10.64~\mathrm{fb}^{-1}$ of $e^+e^-$ collision data taken at center-of-mass energies between 4.237 and 4.699 GeV with the BESIII detector,
we study the leptonic $D^+_s$ decays  using the $e^+e^-\to D^{*+}_{s} D^{*-}_{s}$ process.
The branching fractions of $D_s^+\to\ell^+\nu_{\ell}\,(\ell=\mu,\tau)$
are measured to be
$\mathcal{B}(D_s^+\to\mu^+\nu_\mu)=(\bfmuv)\%$ and
$\mathcal{B}(D_s^+\to\tau^+\nu_\tau)=(\bftauv)\%$, respectively.
The product of the decay constant and Cabibbo-Kobayashi-Maskawa matrix element $|V_{cs}|$ is determined to be $f_{D_s^+}|V_{cs}|=(\mufdsxvcsresult)_{\mu\nu}~\mathrm{MeV}$ and $f_{D_s^+}|V_{cs}|=(\taufdsxvcsresult))_{\tau \nu}~\mathrm{MeV}$, respectively.
Taking the value of $|V_{cs}|$ from a global fit in the Standard Model, we obtain ${f_{D^+_s}}=(\mufdsresult)_{\mu\nu}$\,MeV and ${f_{D^+_s}}=(\taufdsresult)_{\tau \nu}$\,MeV, respectively.  Conversely, taking the value for $f_{D_s^+}$ from the latest lattice quantum chromodynamics calculation, we obtain $|V_{cs}| =(\muvcsresult)_{\mu\nu}$ and
$|V_{cs}| = (\tauvcsresult)_{\tau \nu}$, respectively.
\end{abstract}

\maketitle

\oddsidemargin  -0.2cm
\evensidemargin -0.2cm

\section{Introduction}

Experimental studies of the  decays of the $D^+_s$ are important to understand weak and strong interactions  in the charm decays.
In the Standard Model~(SM), the weak and strong effects in leptonic $D_s^{+}$ decays
can be well-separated. The partial width of the decay
$D_s^+ \to \ell^+\nu_\ell\,(\ell=\mu,\tau)$ is given by~\cite{decayrate}
\begin{equation}
\Gamma_{D_s^+ \to \ell^+\nu_\ell}=
     \frac{G^2_F} {8\pi}
      f^2_{D_s^+} |V_{cs} |^2
      m^2_\ell m_{D_s^+}
    \left (1- \frac{m^2_\ell}{m^2_{D_s^+}}\right )^2,
\label{eq01}
\end{equation}
\noindent
where $G_F$ is the Fermi coupling constant,
$f_{D_s^+}$ is the $D_s^+$ decay constant,
$|V_{cs}|$ is the magnitude of the $c\to s$ Cabibbo-Kobayashi-Maskawa (CKM) matrix element~\cite{pdg2024},
$m_\ell$ is the lepton mass, and $m_{D_s^+}$ is the $D_s^+$ meson mass.
Using the measured branching fraction (BF) of $D_s^+ \to \ell^+\nu_\ell$,
the product of $f_{D_s^+}|V_{cs}|$ can be determined.
By taking the latest $f_{D_s^+}$~\cite{prd98_074512} calculated by lattice quantum
chromodynamics (LQCD), one can determine $|V_{cs}|$, which is an essential input for testing CKM matrix unitarity.
Conversely, taking $|V_{cs}|$ from the SM global fit, one can extract $f_{D_s^+}$,
which is a crucial check of LQCD calculations~\cite{prd98_074512,fds1,fds2,fds3,fds4,fds5}.
In addition, the  BF ratio of $D_s^+ \to \tau^+\nu_\tau$ and $D_s^+ \to \mu^+\nu_\mu$
provides an important  test of $\tau-\mu$ lepton-flavor universality.

In recent years, many studies of $D^+_s\to \ell^+\nu_\ell$ have been performed by the CLEO~\cite{cleo2009,cleo2009a,cleo2009b}, BaBar~\cite{babar2010}, Belle~\cite{belle2013}, and
BESIII~\cite{bes4009,bes3_Ds_muv,bes3_Ds_muv1,bes3_Ds_tauv1,bes3_Ds_tauv2,bes3_Ds_tauv3,xiechenpiv,llcmvv,Ke:2023qzc,Li:2021iwf} experiments. The BESIII Collaboration has reported experimental studies of $D_s^+ \to \ell^+\nu_\ell$ using the $e^+e^-\to D^+_sD^-_s$ and $e^+e^-\to D^{\pm}_sD^{*\mp}_s$ processes. These studies are based on
0.48 fb$^{-1}$ and 7.33 fb$^{-1}$ of $e^+e^-$ collision data taken at the center-of-mass energies($E_{\rm cm}$) of $\sqrt s =$ 4.009  GeV~\cite{bes4009} and $4.128-4.226$ GeV~\cite{bes3_Ds_muv,bes3_Ds_muv1,bes3_Ds_tauv1,bes3_Ds_tauv2,bes3_Ds_tauv3,xiechenpiv,llcmvv}, respectively. The latter ones are the most precise measurements to date.
In this paper,  we perform new measurements of the BFs of $D^+_s\to \mu^+\nu_\mu$ and $D^+_s\to \tau^+\nu_\tau$ via the $e^+e^-\to D^{*+}_{s} D^{*-}_{s}$ process. This analysis utilizes 10.64  fb$^{-1}$ of  $e^+e^-$ collision data collected at center-of-mass energies between $\sqrt s=4.237$ and $4.699$ GeV.  Notably, this is the first time the $e^+e^-\to D_s^{*+}D_s^{*-}$ process has been used to measure $D_s^+$ leptonic decays.
Throughout this paper, charge-conjugation is always implied and $\rho$ denotes the $\rho(770)$ meson.

\section{BESIII detector and MC simulation}
The BESIII detector is a magnetic
spectrometer~\cite{bes3}  operated at the Beijing Electron
Positron Collider (BEPCII)~\cite{Yu:IPAC2016-TUYA01}. The
cylindrical core of the BESIII detector consists of a helium-based
 multilayer drift chamber (${\rm MDC}$), a plastic scintillator time-of-flight
system (${\rm TOF}$), and a CsI(Tl) electromagnetic calorimeter (${\rm EMC}$),
which are all enclosed in a superconducting solenoidal magnet
providing a 1.0~T magnetic field. The solenoid is supported by an
octagonal flux-return yoke with resistive plate counter muon-identifier modules (MUC) interleaved with steel. The acceptance of
charged particles and photons is 93\% over the $4\pi$ solid angle. The
charged-particle momentum resolution at $1~{\rm GeV}/c$ is
$0.5\%$, and the resolution of specific ionization energy loss~(d$E$/d$x$) is $6\%$ for electrons
from Bhabha scattering. The EMC measures photon energies with a
resolution of $2.5\%$ ($5\%$) at $1$~GeV in the barrel (end-cap)
region. The time resolution of the TOF barrel part is 68~ps, while
that of the end-cap part is 110~ps. The end-cap TOF
system was upgraded in 2015 using multi-gap resistive plate chamber technology, providing
a time resolution of 60 ps~\cite{60ps1,60ps2}.
About 74\% of the data used here benefits from this upgrade.  
Details about the design and performance of the BESIII detector are given in Ref.~\cite{bes3}.

Simulated samples produced with the {\sc geant4}-based~\cite{geant4} MC package, which
includes the geometric description of the BESIII detector and the
detector response, are used to determine the detection efficiency
and to estimate the backgrounds. The simulation includes the beam-energy spread and initial-state radiation in the $e^+e^-$
annihilations modeled with the generator {\sc kkmc}~\cite{kkmc}.
An inclusive MC sample with a luminosity of 20 times that of the data is produced at center-of-mass energies between 4.237 and 4.699 GeV. It includes open-charm processes, initial state radiation  production of $\psi(3770)$, $\psi(3686)$ and $J/\psi$, $q\bar q$ $(q=u, d, s)$ continuum processes, along with Bhabha scattering, $\mu^+\mu^-$, $\tau^+\tau^-$, and $\gamma\gamma$ events. In the simulation, the production of open-charm processes directly via $e^+e^-$ annihilations are modeled with the generator {\sc conexc}~\cite{conexc}.
The known decay modes are modeled with {\sc
evtgen}~\cite{evtgen} using the BFs taken from the
Particle Data Group (PDG)~\cite{pdg2024}, and the remaining unknown decays of the charmonium states are
modeled by {\sc lundcharm}~\cite{lundcharm}. Final-state radiation is incorporated using the {\sc photos} package~\cite{photos}.
The input energy-dependent Born cross section for $e^+e^-\to D^{*+}_sD^{*-}_s$ is based on the BESIII measurement \cite{crsDssDss}.

\section{Analysis method}

In the $e^+e^-\to D^{*+}_{s} D^{*-}_{s}$ process, the $D_s$ mesons are produced via $D_s^{*}\to \gamma(\pi^0)D_s$.
We fully reconstruct the transition $\gamma(\pi^0)$ and the $D_s^-$ meson in one of several hadronic decay modes;
successful cases are referred to as single-tag candidates.  
When the single-tag $D_s^{*-}$ and the signal $D_s^{*+}$ decay of interest are simultaneously reconstructed,
we obtain the so-called double-tag candidates.
The BF of the signal  decay is determined by
\begin{equation}
\mathcal B_{\rm sig}=\frac{N_{\rm DT}}{N_{\rm ST} \cdot \bar\epsilon_{{\rm sig}}}.
\label{eq1}
\end{equation}
Here, $N_{\rm DT}$ is the double-tag yield in data; $N_{\rm ST}=\Sigma_{i,j} N_{\rm ST}^{i,j}$ is the total single-tag yield in data summing over the tag mode $i$ and the energy point $j$;
the $\bar \epsilon_{\rm sig}$ is the averaged efficiency of the signal decay, and estimated by $\bar \epsilon_{\rm sig} = \sum_{j} \left[ \sum_{i}\left( \frac{N_{\rm ST}^{i}}{N_{\rm ST}^{j}} \cdot \frac{\epsilon_{\rm DT}^{i}}{\epsilon_{\rm ST}^{i}} \right) \cdot \frac{N_{\rm ST}^{j}}{N_{\rm ST}}\right]$, where $\epsilon_{\rm DT}^{i}$ and $\epsilon_{\rm ST}^{i}$ are the detection efficiencies of the double-tag and single-tag candidates for the $i$-th tag mode, respectively.
The efficiencies include the BFs of the daughter particle decays.

\section{Single-tag $D^{*-}_s$ candidates}

The single-tag $D^{*-}_s$ candidates are formed from the selected $D^-_s$ candidates and a transition $\gamma(\pi^0)$.
The $D^-_s$ candidates are reconstructed from the fourteen hadronic decay modes, including 
$D^-_s\to K^+K^-\pi^-$,
$K^+K^-\pi^-\pi^0$,
$\pi^+\pi^-\pi^-$,
$K^0_SK^-$,
$K^0_SK^-\pi^0$,
$K^-\pi^+\pi^+$, 
$K^0_SK^0_S\pi^-$,
$K^0_SK^+\pi^-\pi^-$,
$K^0_SK^-\pi^+\pi^-$,
$\eta_{\gamma\gamma}\pi^-$,
$\eta_{\pi^0\pi^+\pi^-}\pi^-$,
$\eta^\prime_{\pi^+\pi^-\eta}\pi^-$,
$\eta^\prime_{\gamma\rho^0}\pi^-$, and
$\eta_{\gamma\gamma}\rho^-$. 
Here, the subscripts of the $\eta$ and $\eta^{\prime}$ represent the decay modes used to reconstruct those states.  

All charged tracks must satisfy $|\!\cos\theta|<0.93$, and those not from $K^0_S$ decays
are further required to satisfy $V_{xy}<1$ cm and $|V_{z}|<10$ cm,
where $V_{xy}$ and $V_{z}$ are the distance of the closest approach to the interaction point (IP) in the transverse plane and
along the MDC symmetry axis, respectively, and $\theta$ is the polar angle with respect to the MDC symmetry axis. 
Particle identification (PID) for the charged particles combines measurements of the energy deposited in the MDC~(d$E$/d$x$) and the flight time in the TOF to form likelihoods $\mathcal{L}(h)~(h=p,K,\pi)$ for each hadron $h$ hypothesis.
Charged kaons and pions are identified by comparing the likelihoods for the kaon and pion hypotheses, $\mathcal{L}(K)>\mathcal{L}(\pi)$ and $\mathcal{L}(\pi)>\mathcal{L}(K)$, respectively.

Each $K_{S}^0$ candidate is reconstructed from two oppositely charged tracks satisfying $|V_{z}|<$ 20~cm.
The two charged tracks are assigned
as $\pi^+\pi^-$ without imposing further PID criteria. They are constrained to
originate from a common vertex and are required to have an invariant mass
within $|M_{\pi^{+}\pi^{-}} - m_{K_{S}^{0}}|<$ 12~MeV$/c^{2}$, where
$m_{K_{S}^{0}}$ is the $K^0_{S}$ nominal mass~\cite{pdg2024}. The
decay length of the $K^0_S$ candidate is required to be greater than
twice the vertex resolution away from the IP.

Photon candidates are identified using isolated showers in the EMC.  The deposited energy of each shower must be more than 25~MeV in the barrel region ($|\cos \theta|< 0.80$) and more than 50~MeV in the end cap region ($0.86 <|\cos \theta|< 0.92$).  
To exclude showers that originate from charged tracks,
the angle subtended by the EMC shower and the position of the closest charged track at the EMC
must be greater than 10 degrees as measured from the IP. 
To suppress electronic noise and showers unrelated to the event, the difference between the EMC time and the event start time is required to be within 
[0, 700]\,ns.

The $\pi^0$ and $\eta$ mesons are reconstructed from photon pairs.
To form $\pi^0$ and $\eta$ candidates, the invariant masses of the selected photon pairs are required to be
within the $M_{\gamma\gamma}$ intervals $(0.115,\,0.150)$\,GeV$/c^{2}$ and $(0.500,\,0.570)$\,GeV$/c^{2}$,
respectively.  To improve momentum resolution and suppress background,
a kinematic fit is imposed on each photon pair to constrain their invariant mass to the nominal $\pi^{0}$ or $\eta$ mass~\cite{pdg2024}. The $\chi^2$ of this kinematic fit is required to be less than 20.

The $\eta$ candidates are also formed from $\pi^0\pi^+\pi^-$ combinations with invariant masses 
in the interval $(0.530,\,0.570)~\mathrm{GeV}/c^2$.
The $\eta^\prime$ candidates are formed from the $\eta\pi^+\pi^-$ and $\gamma\rho^0$ combinations
with invariant masses in the intervals $(0.946,\,0.970)$ GeV/$c^2$ and $(0.940,\,0.976)~\mathrm{GeV}/c^2$, respectively.
In addition, the minimum energy
of the $\gamma$ from $\eta'\to\gamma\rho^0$ decays must be greater than 0.1\,GeV.
The $\rho^0$ and $\rho^+$ candidates are reconstructed from the $\pi^+\pi^-$ and $\pi^+\pi^0$
combinations with invariant masses within the common interval $(0.570,\,0.970)~\mathrm{GeV}/c^2$.  

To reject the peaking background events from $D_s^-\to K^0_S(\to \pi^+\pi^-)\pi^-$ and $D_s^-\to K^0_S(\to \pi^+\pi^-)K^-$ in the tag modes of $D_s^-\to \pi^+\pi^-\pi^-$ and $D_s^-\to K^+\pi^+\pi^-$, we require that the invariant mass of any $\pi^+\pi^-$ combination 
	satisfy $|M_{\pi^+\pi^-}-m_{K_S^0}|> 0.03 $  GeV/$c^2$.

\begin{figure*}[htbp]
	\centering
	\includegraphics[width=0.8\textwidth] {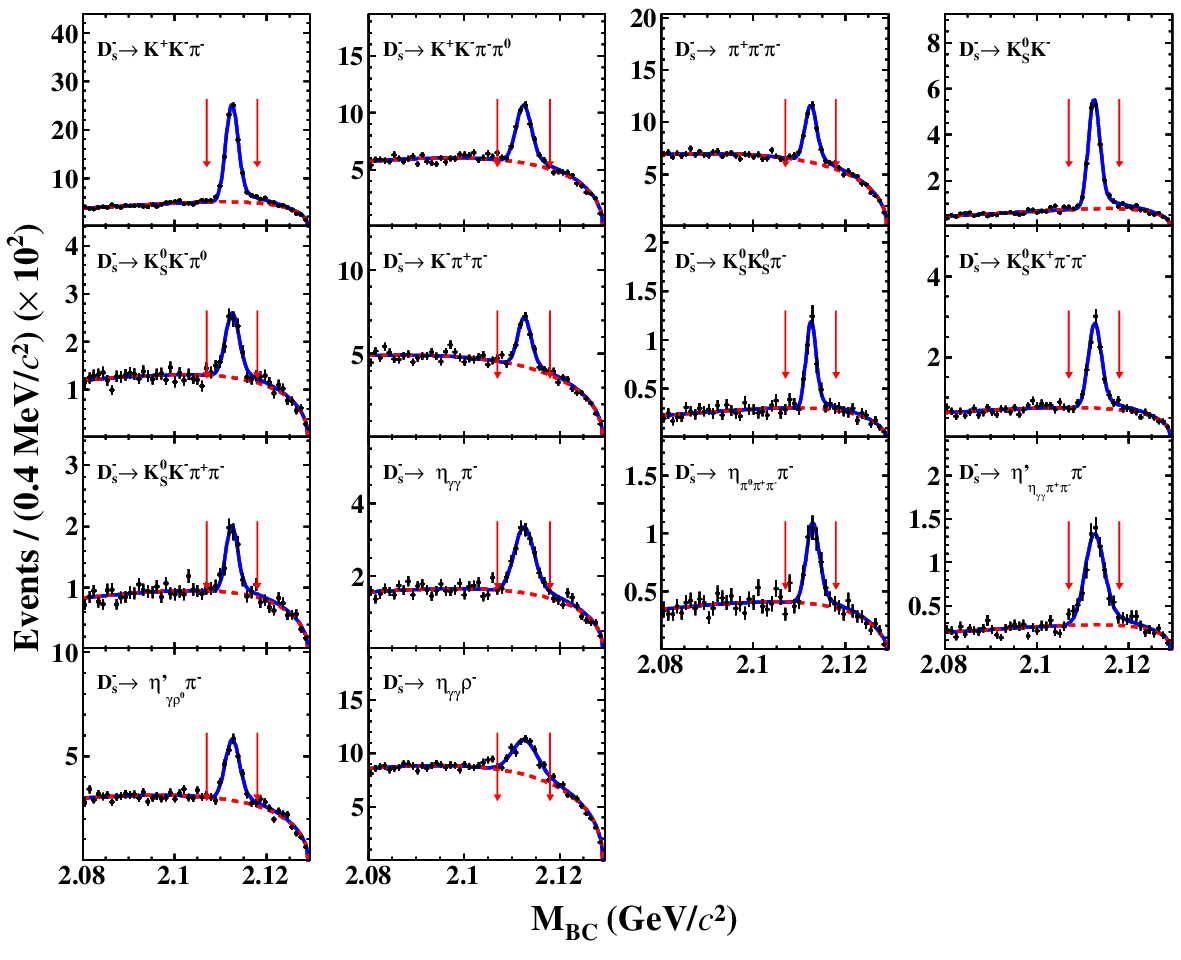}
	\caption{\footnotesize
		Fits to the $M_{\rm BC}$ distributions of the single-tag $D^{*-}_s$ candidates, where the points with error bars are data at 4.260 GeV, the blue solid curves show the best fits, and the red dashed curves show the fitted combinatorial background shapes. shapes. The pairs of arrows denote the $M_{\rm BC}$ signal window. 
		}
	\label{fig:stfit}
\end{figure*}

The tagged $D^-_s$ candidates are required to fall in the $M_{D_s}$ mass windows shown in the second column of Table~\ref{tab:st}, following Refs.~\cite{panx1,panx2}. To further  distinguish the single-tag $D^{*-}_s$ from combinatorial background, we use two kinematic variables:
the energy difference defined as
\begin{equation}
\Delta E =  E_{\rm tag} - E_{\rm beam},
\label{eq:deltaE}
\end{equation}
and the beam-constrained mass defined as
\begin{equation}
M_{\rm BC}=\sqrt{E^2_{\rm beam}/c^4-|\vec{p}_{\rm tag}|^2/c^2},
\label{eq:mBC}
\end{equation}
where $E_{\rm beam}$ is the beam energy, and $E_{\rm tag}$ and
$\vec{p}_{\rm tag}$ are the energy and momentum of the single-tag $D^{*-}_s$ candidate in the rest frame of the initial $e^+e^-$ beams.

The $\Delta E$ values for each tag must be within the ranges listed in Table~\ref{tab:st}.  
If multiple $\gamma/\pi^0$ or $D_s$ combinations remain, 
we keep only the candidates with the minimum $|\Delta E|$
for each tag mode and each $D_s^*$ charge from a given event. 
For each tag mode, the single-tag yield is obtained by a fit to
the corresponding $M_{\rm BC}$ spectrum.
The signal is described by the simulated shape convolved with a Gaussian function
representing the difference in resolution between data and simulation.
Because the endpoint above 4.450 GeV is far away from the nominal $D^*_s$ mass, the data at all energy points above 4.450 GeV are combined into one data set. For the data sets taken below and above 4.450 GeV, the non-peaking background shapes are modeled by an ARGUS function~\cite{argus}  and a second order Chebychev polynomial function, respectively, which have been validated by analyzing the inclusive MC sample.
As an example, the $M_{\rm BC}$ distributions of accepted single-tag candidates for various tag modes and the fit results at 4.260 GeV are shown in Fig.~\ref{fig:stfit}.
The candidates in the signal regions, denoted as the red arrows in each sub-figure, are kept for further analysis.
The resulting single-tag yields ($N_{\rm ST}^{i}$) for various tag modes and the corresponding single-tag efficiencies ($\epsilon_{\rm ST}^{i}$) at 4.260 GeV are
shown in  Table~\ref{tab:st}.
Information for data from other energies is given in Appendix A.  
The total single-tag yields at the different energy points are summarized in Table~\ref{tab:sigeff}.

\begin{table*}[htbp]
	\caption{The $M_{D^-_s}$ requirements, $\Delta E$ requirements, single-tag yields ($N_{\rm ST}$), single-tag efficiencies ($\epsilon_{\rm ST}$), and various double-tag efficiencies for each tag mode at 4.260~GeV. The $\epsilon^{\mu_a\nu}_{\rm DT}$ and $\epsilon^{\mu_b\nu}_{\rm DT}$ correspond to the double-tag efficiencies
		for $D^+_s\to \mu^+_a\nu_\mu$ and $D^+_s\to \mu^+_b\nu_\mu$, while $\epsilon^{\tau_e\nu}_{\rm DT}$, $\epsilon^{\tau_\mu\nu}_{\rm DT}$, $\epsilon^{\tau_\pi\nu}_{\rm DT}$, and $\epsilon^{\tau_\rho\nu}_{\rm DT}$ correspond to the double-tag efficiencies for
		$D^{+}_{s} \to \tau^+_e\nu_{\tau}$,
		$D^{+}_{s} \to \tau^+_\mu\nu_{\tau}$,
		$D^{+}_{s} \to \tau^+_\pi\nu_{\tau}$, and
		$D^{+}_{s} \to \tau^+_\rho\nu_{\tau}$, respectively. The
		Muon candidates identified with (without) MUC information are denoted by $\mu_a$ ($\mu_b$). 
                The efficiencies include the
		BFs of $D_s^{*+}$ and $\tau^+$ decays. The uncertainties are statistical only.
		\label{tab:ST}}
	\centering
	\label{tab:st}
	\resizebox{1.0\textwidth}{!}{
\begin{tabular}{ l l l   r@{} lr@{}  lr@{} lr@{} lr@{} lr@{} lr@{} lr@{} lr@{}l }
	\hline\hline
\makecell{$D^-_s$ tag \\mode} &\makecell{$M_{D^-_s}$\\(GeV/$c^2$)}&\makecell{$\Delta E$\\(MeV)} &\multicolumn{2}{c}{$N_{\rm ST}$} &\multicolumn{2}{c}{\makecell{$\epsilon_{\rm ST}$\\(\%)}}
&\multicolumn{2}{c}{\makecell{$\epsilon^{\mu_a\nu}_{\rm DT}$\\(\%)}}&\multicolumn{2}{c}{\makecell{$\epsilon^{\mu_b\nu}_{\rm DT}$\\(\%)}}&\multicolumn{2}{c}{\makecell{$\epsilon^{\tau_e\nu}_{\rm DT}$\\(\%)}}&\multicolumn{2}{c}{\makecell{$\epsilon^{\tau_\mu\nu}_{\rm DT}$\\(\%)}}&\multicolumn{2}{c}{\makecell{$\epsilon^{\tau_\pi\nu}_{\rm DT}$\\(\%)}}&\multicolumn{2}{c}{\makecell{$\epsilon^{\tau_\rho\nu}_{\rm DT}$\\(\%)}} \\
	\hline
  $K^{+} K^{-}\pi^{-}$                              &$(1.950,1.986)$&$(-31     ,26)$&$7454$&$\pm125$&$19.67   $&$\pm0.07$&$14.43   $&$\pm0.07$&$16.89   $&$\pm0.09$&$3.54    $&$\pm0.02$&$1.53    $&$\pm0.01$&$1.62    $&$\pm0.01$&$1.07    $&$\pm0.01$\\
  $K^{+} K^{-}\pi^{-}\pi^{0}$                       &$(1.947,1.982)$&$(-38     ,29)$&$2186$&$\pm108$&$5.15    $&$\pm0.05$&$5.08    $&$\pm0.06$&$5.53    $&$\pm0.06$&$1.21    $&$\pm0.01$&$0.55    $&$\pm0.01$&$0.52    $&$\pm0.01$&$0.28    $&$\pm0.01$\\
  $\pi^{-}\pi^{+}\pi^{-}$                           &$(1.952,1.984)$&$(-34     ,28)$&$1929$&$\pm99$&$25.73   $&$\pm0.26$&$18.18   $&$\pm0.08$&$24.60   $&$\pm0.10$&$4.32    $&$\pm0.02$&$1.84    $&$\pm0.01$&$2.35    $&$\pm0.01$&$1.76    $&$\pm0.02$\\
  $K_S^{0} K^{-}$                                   &$(1.948,1.991)$&$(-33     ,30)$&$1649$&$\pm53$&$22.97   $&$\pm0.16$ &$16.44   $&$\pm0.09$&$21.80   $&$\pm0.11$&$3.87    $&$\pm0.02$&$1.68    $&$\pm0.01$&$2.07    $&$\pm0.01$&$1.53    $&$\pm0.02$\\$
  K_S^{0} K^{-}\pi^{0}$                            &$(1.946,1.987)$&$(-40     ,31)$&$554$&$\pm50$&$7.51    $&$\pm0.14$&$6.45    $&$\pm0.09$&$8.73    $&$\pm0.11$&$1.54    $&$\pm0.02$&$0.68    $&$\pm0.01$&$0.84    $&$\pm0.01$&$0.53    $&$\pm0.01$\\
  $K^{-}\pi^{-}\pi^{+}$                             &$(1.953,1.983)$&$(-33     ,28)$&$1112$&$\pm83$&$23.47   $&$\pm0.40$&$16.15   $&$\pm0.17$&$21.19   $&$\pm0.20$&$3.83    $&$\pm0.03$&$1.66    $&$\pm0.02$&$1.98    $&$\pm0.02$&$1.46    $&$\pm0.03$\\
  $K_S^{0} K_S^{0}\pi^{-}$                          &$(1.951,1.986)$&$(-32     ,28)$&$266$&$\pm22$&$11.07   $&$\pm0.22$&$7.96    $&$\pm0.14$&$10.84   $&$\pm0.18$&$1.94    $&$\pm0.03$&$0.83    $&$\pm0.02$&$1.03    $&$\pm0.02$&$0.72    $&$\pm0.02$\\
  $K_S^{0} K^{+}\pi^{-}\pi^{-}$                     &$(1.953,1.983)$&$(-31     ,26)$&$808$&$\pm45$&$10.21   $&$\pm0.12$&$7.46    $&$\pm0.12$&$9.30    $&$\pm0.14$&$1.80    $&$\pm0.02$&$0.78    $&$\pm0.02$&$0.88    $&$\pm0.01$&$0.52    $&$\pm0.02$\\
  $K_S^{0} K^{-}\pi^{+}\pi^{-}$                     &$(1.958,1.980)$&$(-31     ,26)$&$390$&$\pm40$&$9.66    $&$\pm0.20$&$6.70    $&$\pm0.06$&$8.51    $&$\pm0.07$&$1.64    $&$\pm0.01$&$0.73    $&$\pm0.01$&$0.79    $&$\pm0.01$&$0.50    $&$\pm0.01$\\
  $\eta_{\gamma\gamma}\pi^{-}$                      &$(1.930,2.000)$&$(-52     ,43)$&$983$&$\pm69$&$19.33   $&$\pm0.29$&$15.89   $&$\pm0.08$&$22.99   $&$\pm0.09$&$3.81    $&$\pm0.02$&$1.63    $&$\pm0.01$&$2.24    $&$\pm0.01$&$1.69    $&$\pm0.01$\\
  $\eta_{\pi^{+}\pi^{-}\pi^{0}}\pi^{-}$             &$(1.941,1.990)$&$(-43     ,34)$&$269$&$\pm29$&$11.22   $&$\pm0.28$&$8.47    $&$\pm0.06$&$11.92   $&$\pm0.07$&$2.02    $&$\pm0.01$&$0.90    $&$\pm0.01$&$1.14    $&$\pm0.01$&$0.79    $&$\pm0.01$\\
  $\eta\prime_{\pi^{+}\pi^{-}\eta} \pi^{-}$         &$(1.940,1.996)$&$(-40     ,34)$&$575$&$\pm40$&$11.16   $&$\pm0.18$&$8.60    $&$\pm0.06$&$12.39   $&$\pm0.07$&$2.05    $&$\pm0.01$&$0.91    $&$\pm0.01$&$1.18    $&$\pm0.01$&$0.85    $&$\pm0.01$\\
  $\eta\prime_{\gamma\rho^{0}} \pi^{-}$             &$(1.938,1.992)$&$(-43     ,33)$&$1233$&$\pm75$&$14.00   $&$\pm0.19$&$11.04   $&$\pm0.07$&$15.46   $&$\pm0.08$&$2.68    $&$\pm0.01$&$1.17    $&$\pm0.01$&$1.49    $&$\pm0.01$&$1.00    $&$\pm0.01$\\
  $\eta_{\gamma\gamma}\rho^{-}$                     &$(1.920,2.006)$&$(-66     ,49)$&$2142$&$\pm191$&$8.07    $&$\pm0.13$&$7.64    $&$\pm0.06$&$11.59   $&$\pm0.07$&$1.83    $&$\pm0.01$&$0.80    $&$\pm0.01$&$1.13    $&$\pm0.01$&$0.80    $&$\pm0.01$\\
\hline\hline
\end{tabular}
}
\end{table*}

\begin{table}[htbp]\centering
	\caption{The integrated luminosities, ${\mathcal L}$, the $M_{\rm BC}$ requirements, and the single-tag yields in data, $N_{\rm ST}$, for various energy points. The uncertainties are statistical only.}
	\label{tab:sigeff}.
		\begin{tabular}{lcc r@{}l}
			\hline
			\hline
			$E_{\rm cm}$ (GeV)&${\mathcal L}$ (pb$^{-1}$) &$M_{\rm BC}$ (GeV/$c^2$) &\multicolumn{2}{c}{\makecell{ $N_{\rm ST}$ }}\\
			\hline
		4.237       &530.3&$(2.107,2.117)$&$6477$&$\pm163$    \\
		4.246       &593.9&$(2.107,2.118)$&$11944$&$\pm246$    \\
		4.260       &828.4&$(2.107,2.118)$&$21550$&$\pm320$    \\
		4.270       &531.1&$(2.107,2.118)$&$13319$&$\pm244$   \\
		4.280       &175.7&$(2.106,2.119)$&$4063$&$\pm152$    \\
		4.290       &502.4&$(2.106,2.119)$&$9316$&$\pm221$    \\
		4.310-4.315       &546.3&$(2.106,2.119)$&$5758$&$\pm228$    \\
		4.400       &507.8&$(2.106,2.119)$&$1855$&$\pm87$     \\
		4.420       &1090.7&$(2.106,2.121)$&$14890$&$\pm443$   \\
		4.440       &569.9&$(2.106,2.121)$&$9699$&$\pm443$    \\
		4.470-4.699 &4768.3&$(2.104,2.123)$&$25156$&$\pm762$   \\
		Sum   & \multicolumn{2}{c}{ }&$124027$&$  \pm1121$\\
			
			\hline\hline
		\end{tabular}
	\end{table}

\section{Analysis of leptonic $D^+_s$ decays}

\subsection{Selection of double-tag events}

The candidates for $D^{*+}_s$ containing a leptonic $D^+_s$ decay are reconstructed from the unused 
showers and tracks remaining after the single-tag selection.
We require that only one additional track remains after the tag reconstruction. To further suppress hadronic background, we require that there is no extra  charged track in each candidate event: $N_{\rm extra}^{\rm charge}$= 0.
The transition $\gamma(\pi^0)$ is selected with the same selection criteria as the tag side.

Throughout this paper, the
$\mu^+_a$  denote the $\mu^+$ candidates identified with MUC information, 
the $\mu_b$ denote the $\mu^+$ candidates identified without MUC information;
and the $\tau^+_e$, $\tau^+_\mu$, $\tau^+_\pi$, and $\tau^+_\rho$ denote
the $\tau^+$ candidates reconstructed via
$\tau^+ \to e^+\nu_e \bar \nu_{\tau}$,
$\tau^+ \to \mu^+\nu_\mu \bar \nu_{\tau}$,
$\tau^+ \to \pi^+\bar\nu_\tau$, and
$\tau^+ \to \rho^+\bar\nu_\tau$, respectively. In the selection of the candidates for $D^+_s\to \tau^+_\pi\nu_\tau$ and
$D^+_s\to \tau^+_\rho\nu_\tau$, the $\pi^+$ and $\rho^+$ candidates
are selected with the same selection criteria as those used on the tag side.
To select the candidates for $D^+_s\to \tau^+_e\nu_\tau$,
the positron PID uses the measured information in the MDC, TOF, and EMC. 
The combined likelihoods ($\mathcal{L}'$)
under the positron, pion, and kaon hypotheses are obtained.
Positron candidates are required to satisfy $\mathcal{L}'(e)>0.001$ and $\mathcal{L}'(e)/(\mathcal{L}'(e)+\mathcal{L}'(\pi)+\mathcal{L}'(K))>0.8$.
To select the candidates for $D^+_s\to \tau^+_\mu\nu_\tau$, the muon candidate is required to have a
deposited energy in the EMC within $(0.1,\,0.3)$\,GeV.
It must also satisfy a requirement on the hit depth, $d_{\mu^{+}}$, in the muon counter 
which depends on both $|\cos\theta_{\mu^{+}}|$ and $p_{\mu^+}$.  
These requirements on $d_{\mu^+}$ are shown in Table~\ref{tab:muonid}.
The selected $D^+_s\to \tau^+_{\pi}\nu_{\tau}$ candidates offer an opportunity to determine
the BF of $D^+_s\to \mu^+_b\nu_\mu$ as a cross check.

\begin{table}[hbtp]
	\begin{center}
		\caption{The $\cos\theta_{\mu^+}$ and $p_{\mu^+}$ dependent requirements
			on $d_{\mu^+}$ for muon candidates.
			}
		\begin{tabular}{ccc} \hline\hline
			$|\cos\theta_{\mu^+}|$ & $p_{\mu^+}$ (GeV/$c$) & $d_{\mu^+}$ (cm) \\ \hline
			&$p_{\mu^+}\le0.88$      &$d_{\mu^+}>17.0$        \\
			$$(0.00, 0.20)$$ \;\; &$0.88<p_{\mu^+}<1.04$ \;\; &$d_{\mu^+}>100.0 p_{\mu^+}-71.0$ \; \\
			&$p_{\mu^+}\ge1.04$      &$d_{\mu^+}>33.0$ \\ \hline
			&$p_{\mu^+}\le0.91$      &$d_{\mu^+}>17.0$        \\
			$$(0.20, 0.40)$$&$0.91<p_{\mu^+}<1.07$&$d_{\mu^+}>100.0 p_{\mu^+}-74.0$ \\
			&$p_{\mu^+}\ge1.07$      &$d_{\mu^+}>33.0$ \\ \hline
			&$p_{\mu^+}\le0.94$      &$d_{\mu^+}>17.0$        \\
			$$(0.40, 0.60)$$&$0.94<p_{\mu^+}<1.10$&$d_{\mu^+}>100.0 p_{\mu^+}-77.0$ \\
			&$p_{\mu^+}\ge1.10$      &$d_{\mu^+}>33.0$ \\ \hline
			$$(0.60, 0.80)$$&         ...         &$d_{\mu^+}>17.0$ \\ \hline
			$$(0.80, 0.93)$$&          ...        &$d_{\mu^+}>17.0$ \\ \hline\hline
		\end{tabular}
		\label{tab:muonid}
	\end{center}
\end{table}

Information concerning the undetectable neutrino(s) is inferred by the kinematic quantity
$M^2_{\mathrm{miss}}\equiv E^2_{\mathrm{miss}}/c^4-|\vec{p}_{\mathrm{miss}}|^2/c^2$,
where $E_{\mathrm{miss}}$ and $\vec{p}_{\mathrm{miss}}$ are the missing energy and momentum
of the neutrino(s) candidate, respectively,
calculated by $E_{\mathrm{miss}}\equiv E_{\mathrm{beam}}- E_k$
and $\vec{p}_{\mathrm{miss}}\equiv-\vec{p}_{D_s^{*-}}-\vec{p}_{\gamma(\pi^0)}-\vec{p}_k$
in the $e^+e^-$ center-of-mass frame. The index $k$ denotes the $e^+$, $\mu^+$,  $\pi^+$, or $\rho^+$  of the signal candidate,
and $E_k$ and $\vec{p}_k$ are its energy and momentum, respectively. 
For $D_s^+\to\mu^+_a\nu_{\mu}$, to improve the $M_{\rm miss}^2$ resolution,
we perform a kinematic fit that constrains the masses of all possible particles and the missing neutrino combinations to the known mass of $D^{*-}_s$
and $D^{*+}_s$.
If there is more than one combination, the one with the minimum $\chi^2$ values of the kinematic fit
is retained for further analysis.
The variable $E_{\rm sum}^{\rm extra\,\gamma}$  is defined as the total energy of
good showers in the EMC including the transition $\gamma(\pi^0)$  used to reconstruct the signal
$D_s^{*+}$, but excluding those used in the tag side.
We also remove bremsstrahlung photons candidates, defined as showers reconstructed
within 10 degrees of the initial positron track direction.  

\begin{figure}[htbp]
	\centering
	\includegraphics[width=1.0\linewidth]{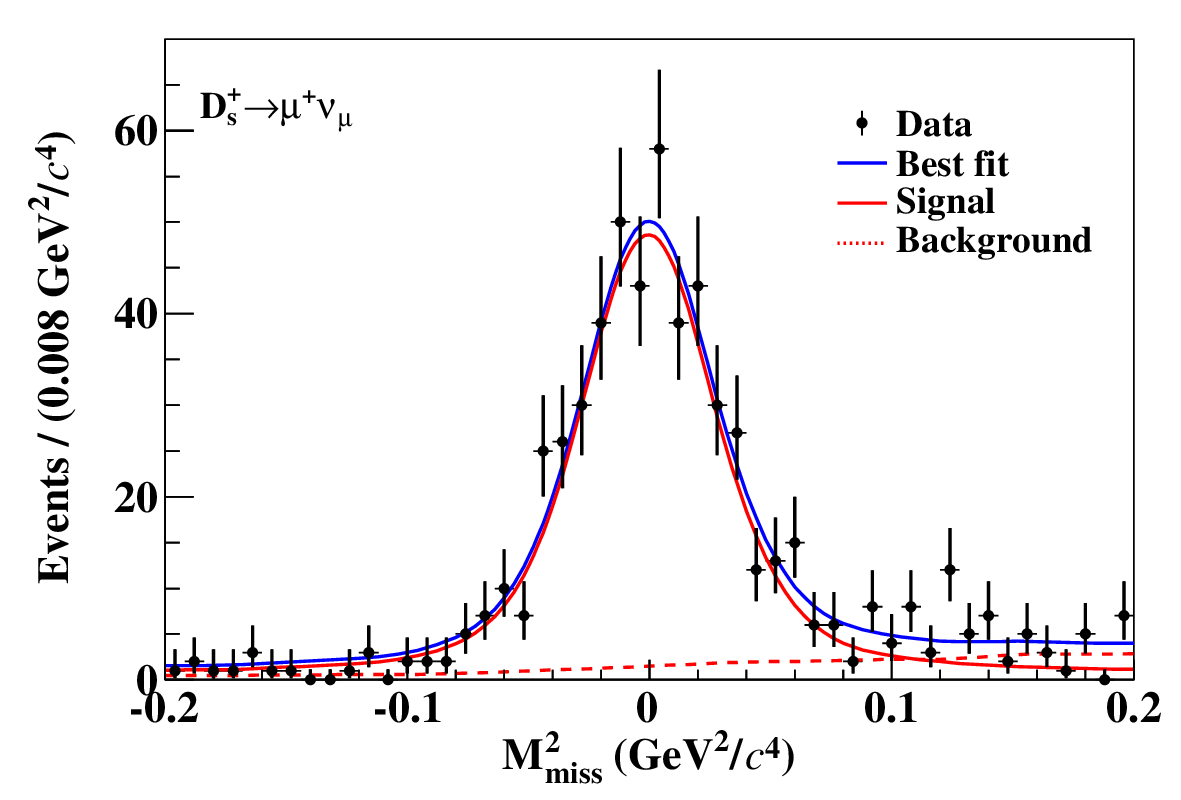}
	\caption{Fit to the $\mm$ distribution of the candidate events for $\muv$  with the $\mu^+$ depth requirement. 
		The points with error bars represent data combined from all energy points. The blue solid curve denotes the total fit.
		The red solid curves and red dashed curves show the fitted signal 
		and combinatorial background shapes.
\label{fig:Dstomunu}}
\end{figure}

\begin{figure*}[htbp]
	\centering
	\includegraphics[width=0.8\linewidth]{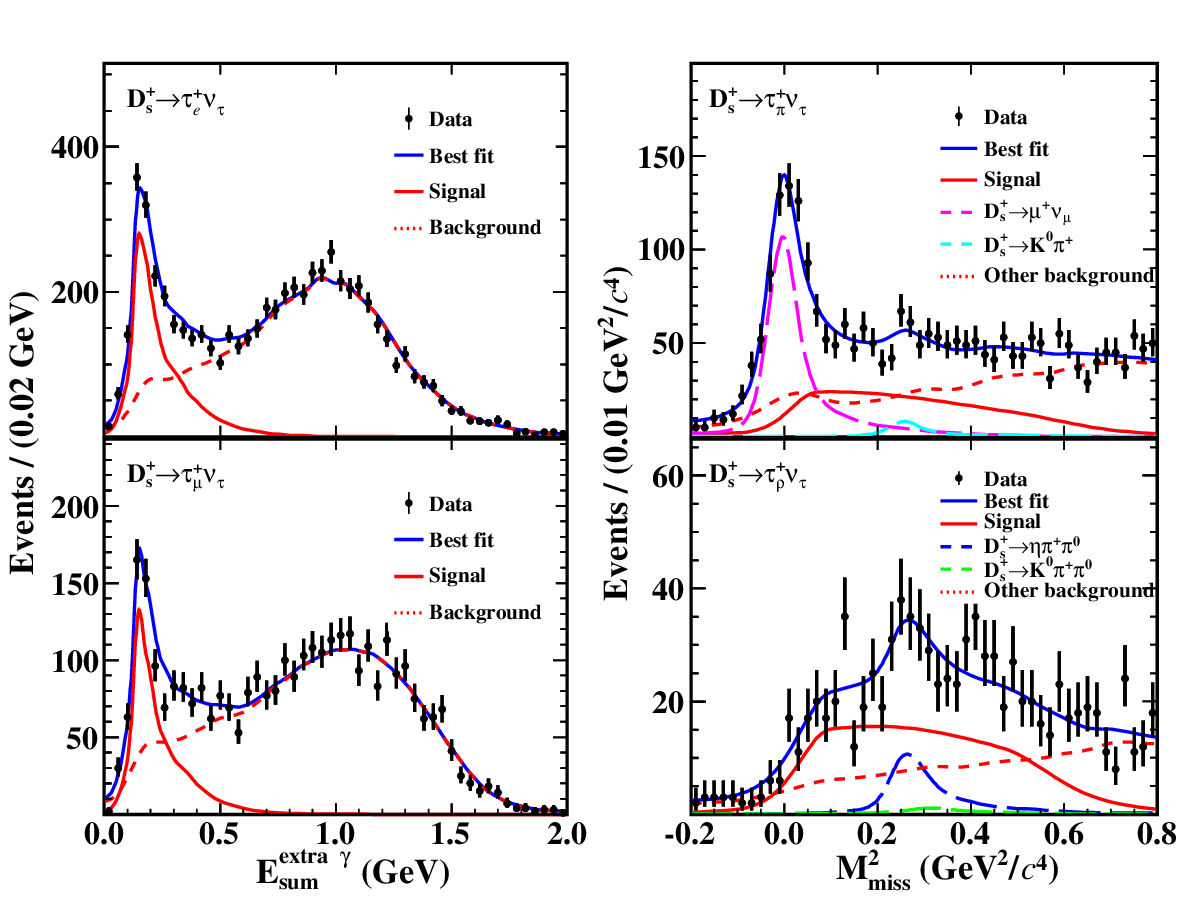}
	\caption{Simultaneous fit to the
$E_{\rm sum}^{\rm extra~\gamma}$ distributions (left) of
 $D^{+}_{s} \to \tau^+_e \nu_{\tau}$ and $D^{+}_{s} \to \tau^+_\mu \nu_{\tau}$
as well as the $\mm$ distributions (right) of $D^{+}_{s} \to \tau^+_\pi\nu_{\tau}$ and $D^{+}_{s} \to \tau^+_\rho\nu_{\tau}$.
		The points with error bars represent data combined from all energy points. The blue solid curve denotes the total fit.
		The red solid curves and red dashed curves show
		the fitted combinatorial background shapes. The pink dashed, cyan dashed  blue dashed, and green dashed are the backgrounds from $D_s^+\to \mu^+\nu_{\mu^+}$, $D_s^+\to K^0\pi^+$, $D_s^+\to\eta\pi^+\pi^0$ and $D_s^+\to K^0\pi^+\pi^0$, respectively.
\label{fig:Dstotaunu}}
\end{figure*}

 Figure \ref{fig:Dstomunu} shows  the $M^{2}_{\rm miss}$ distribution of the candidate events for $D^+_s\to \mu_{a}^+\nu_\mu$.
The left two plots of Fig.~\ref{fig:Dstotaunu} show  the $E_{\rm sum}^{\rm extra\,\gamma}$ distributions of
the candidate events for $D^+_s\to \tau^+_e\nu_\tau$ and $D^+_s\to \tau^+_\mu\nu_\tau$,
and the right two plots of Fig.~\ref{fig:Dstotaunu} show the $M^{2}_{\rm miss}$ distributions of the candidate events for $D^+_s\to \tau^+_\pi\nu_\tau$ and $D^+_s\to \tau^+_\rho\nu_\tau$. 
Different signal variables are chosen for the different $\tau^+$ modes mainly to achieve better separation between 
signal and background. 
There  are no prominent peaking backgrounds for $D^+_s\to \tau^+_e\nu_\tau$ and $D^+_s\to \tau^+_\mu\nu_\tau$, while the main peaking backgrounds are from $D^+_s\to\mu^+\nu_{\mu}$ and $D^+_s\to K^0\pi^+$ for $D^+_s\to \tau^+_\pi\nu_\tau$, and the main peaking  backgrounds are from  $D^+_s\to\eta\pi^+\pi^0$ and $D^+_s\to K^0\pi^+\pi^0$ for $D^+_s\to \tau^+_\rho\nu_\tau$.
These distributions are obtained from the combined data from all energy points.

\subsection{Detection efficiencies}

The detection efficiencies, $\epsilon_{\rm ST}^{ij}$ and  $\epsilon_{\rm DT}^{ij}$,  are estimated by analyzing 
MC samples with the relevant combinations of events and appropriate relative BFs.  
The single-tag MC sample is generated with inclusive $D^{*-}_s$ decays and $D^-_s$ decays to the single modes,
while the double-tag has $D^{*-}_s$ decays to signal modes and $D^-_s$ decays to the tag modes.  
and $D^{*+}_s\to \rm anything$ with $D^+_s$ decays to signal modes.
As an example,  the double-tag efficiencies  obtained for each signal decay at 4.260 GeV are shown in Table~\ref{tab:ST}.
For a given tag mode $i$ and energy point $j$, the effective signal efficiencies, 
$\epsilon_{\rm sig}^{ij}$, of each signal decay are obtained by
dividing the $\epsilon_{\rm DT}^{ij}$ by $\epsilon_{\rm ST}^{ij}$.
For each signal decay, the averaged signal efficiencies, $\bar \epsilon_{\rm sig}$, are obtained by weighting them 
by the relative single-tag yields $N_{\rm ST}^{ij}$; the results are shown in Table~\ref{tab:branching fractions _data}.

\subsection{Branching fractions without lepton universality constraint}

When using the MUC information to identify muon candidates, the background level of $D_s^+\to\mu_a^+\nu_\mu$ is very low.
In this case, the signal yield of $D_s^+\to\mu_a^+\nu_\mu$ is obtained from the unbinned maximum likelihood fit to the $M^2_{\rm miss}$ distribution, as shown in Fig.~\ref{fig:Dstomunu}.
In this fit, the signal and background shapes are modeled by the simulated shapes.
We obtain a signal yield of $507\pm26$. 
Combining this yield and detection, the efficiency, and the single-tag yield,
we use Eq.~\ref{eq1} to obtain the BF of $D_s^+\to\mu_a^+\nu_\mu$ result: 
\begin{equation}\mathcal{B}_{D_s^+\to\mu^+_a\nu_\mu}=(\bfmuv)\%.\nonumber
\end{equation}
The systematic uncertainties in the BF measurement are discussed in the following section.

To extract the BF of $D^+_s\to \tau^+\nu_\tau$, 
we perform a simultaneous fit to the
$E_{\rm sum}^{\rm extra~\gamma}$ distributions of
candidates for $D^{+}_{s} \to\tau^+_\ell \tau_\tau$
($\ell=e$ or $\mu$) and  the $\mm$ distributions of candidates for $D^{+}_{s} \to \tau^+_h \tau_\tau$,( $h=\pi^+$ or $\rho^+$), 
as shown in Fig.~\ref{fig:Dstotaunu}. 
 There is large component of $D^+_s\to \mu_b^+\nu_\mu$ in the accepted candidates $D^{+}_{s} \to \tau^+_\pi\nu_\tau$  due to the poorer separation muons and pions 
when using only the $dE/dx$, TOF, and EMC information. 
The signal shapes of $D^+_s\to \mu_a^+\nu_\mu$, $D^+_s\to \mu_b^+\nu_\mu$, $\Dstotaunue$,
and  $\Dstotaunum$ are modeled by the individual simulated shapes for each decay mode.
The signal shapes of the $\Dstotaunup$ and  $\Dstotaunur$  are described by
a sum of two bifurcated-Gaussian functions, whose parameters are determined
from the fits to the signal MC events and are fixed in the simultaneous fit.
The  background components for each signal decay are modeled with shapes derived from
the relevant MC simulation.   
The four $\tau^+$ decay modes are constrained to have a common BF for $D^+_s\to \tau^+\nu_\tau$, 
taking into account different efficiencies and daughter particle decay BFs. 
From this simultaneous fit, we obtain 
\begin{equation}\mathcal{B}_{D_s^+\to\mu^+_b\nu_\mu}=(0.493\pm0.029_{\rm stat}\pm0.020_{\rm syst})\%\nonumber\end{equation}
and
\begin{equation}\mathcal{B}_{D_s^+\to\tau^+\nu_\tau}=(\bftauv)\%,\nonumber\end{equation}
which correspond to total signal yields of $579\pm34$ and $2845\pm83$, respectively. 
The systematic uncertainties in the BF measurements are discussed in the following section.

\subsection{Branching fractions with lepton universality constraint}

The ratio of decay widths between $D^+_s\to \tau^+\nu_\tau$ and $D^+_s\to \mu^+\nu_\mu$ is given by 
\begin{equation}\label{eq:theR}
	R = \frac{\Gamma_{D^+_s\to\tau^+\nu_\tau}}{\Gamma_{D^+_s\to\mu^+\nu_\mu}}
	= \frac{m^2_\tau(1-\frac{m^2_\tau}{m^2_{D_s}})^2}{m^2_\mu(1-\frac{m^2_\mu}{m^2_{D_s}})^2}.
\end{equation}
\noindent In this ratio, $f_{D^+_s}$ and $|V_{cs}|$ cancel, we can obtain  a very precise SM prediction of $R = 9.75\pm0.01$, with the $\mu^+$ and $\tau^+$ masses from the PDG~\cite{pdg2024}.
To improve the precision of the measured $f_{D_s^+}|V_{cs}|$, we  have  examined  the BFs of $D^+_s\to \mu^+\nu_\mu$ and $D^+_s\to \tau^+\nu_\tau$ after further constraining the ratio ${\mathcal B}(D^+_s\to \tau^+\nu_\tau)/{\mathcal B}(D^+_s\to \mu^+\nu_\mu)$ to be $9.75$ based on the SM prediction,   in the fits to individual distributions in Fig. \ref{fig:Dstotaunu}. From this constrained fit, the BFs of $D_s^+\to\mu^+\nu_\mu$ and $D_s^+\to\tau^+\nu_\tau$ are
obtained to be
\begin{equation}
\mathcal{B}_{D_s^+\to\mu^+\nu_\mu}^{\rm SM}=(\Rbfmuv)\%,\nonumber\end{equation}
 and
 \begin{equation}
\mathcal{B}_{D_s^+\to\tau^+\nu_\tau}^{\rm SM}=(\Rbftauv)\%,
\nonumber\end{equation}
respectively. These correspond to signal yields of $D_s^+\to\mu^+\nu_\mu$ and $D_s^+\to\tau^+\nu_\tau$ to
be $641\pm16$ and $2754\pm69$. 
The systematic uncertainties in the BF measurements are discussed in the following section.

\begin{table*}[htbp]
	\centering
	\caption{
		The signal yields, the effective signal efficiencies ($\bar \epsilon_{\rm sig}$), and the obtained BFs.
		The efficiencies include the BFs  of all sub-resonant decays.
		The first uncertainties are statistical and the second systematic. The high effective signal efficiency for $D^+_s\to \mu^+\nu_\mu$ is mainly due to that the tag environments in the signal MC sample is very different with the inclusive MC sample, where the single-tag efficiency is affected much due to low momentum photon(s) and pions.
		\label{tab:branching fractions _data}
	}
	\begin{tabular}{l|c|cc|cc}
		\hline
		\hline
		 \multirow{2}{*}{Signal decay} &\multirow{2}{*}{$\bar \epsilon_{\rm sig}$ (\%)}&\multicolumn{2}{c|}{No lepton universality constraint}&\multicolumn{2}{c}{ Lepton universality constraint}\\
	    &                 & $N_{\rm DT}$     & $\mathcal B$ (\%)  & $N^{\rm SM}_{\rm DT}$& $\mathcal B^{\rm SM}$ (\%)\\ \hline
	
	$D^{+}_{s} \to \tau^+_e\nu_{\tau}$         &$7.81    \pm0.02$ & \multirow{4}{*}{$2845 \pm 83$}     & \multirow{4}{*}{$5.60 \pm 0.16\pm0.20$}
	& \multirow{4}{*}{$2754 \pm 69$}&\multirow{4}{*}{$5.39 \pm 0.14\pm0.20$}\\
	$D^{+}_{s} \to \tau^+_\mu\nu_{\tau}$       &$18.57   \pm0.04$  &        &                          &   &                                    \\
	$D^{+}_{s} \to \tau^+_\pi\nu_{\tau}$                 &$8.93    \pm0.02$ &       &&&\\
	$D^{+}_{s} \to \tau^+_\rho\nu_{\tau}$                  &$6.11    \pm0.02$ &       &&                          &                                    \\
	\hline
	$D_{s}^{+}\to \mu^{+}_b\nu_{\mu}$                                           &$94.76   \pm0.20$ & $579\pm34$ &$0.491   \pm0.029\pm0.020$&$641\pm16$&$0.553   \pm0.014\pm0.021$\\
	\hline
	$D_{s}^{+}\to \mu^{+}_a\nu_{\mu}$                                             &$74.67   \pm0.16$ &$507\pm26$   &$0.547  \pm0.026\pm0.016$&...&...\\
		\hline
		\hline
	\end{tabular}
\end{table*}

\section{Systematic uncertainties}

The systematic uncertainties associated with the single-tag selection cancel. Several other sources of systematic uncertainties are estimated for the BFs measurements and described below.
\subsection{Individual systematic sources}
\subsubsection{Single-tag yield}

To estimate the systematic uncertainty in the fitted single-tag yield, we perform fits to the $M_{\rm BC}$ distributions of the data and inclusive MC sample with alternative signal and background shapes.
The nominal signal shape is the simulated shape convolved with a Gaussian function.
 An alternative signal shape is obtained after requiring that the angle between each reconstructed and generated track is less than 20 degrees. The background shape is changed to a third-order Chebychev polynomial. The relative differences of the ST yields between
data and the inclusive MC sample from these two variations are added in quadrature assigned as the systematic uncertainty. 
In addition, the uncertainty due to the background fluctuation of the fitted single-tag yield is considered as an additional systematic uncertainty.
The quadrature sum of these two items, 1.88\%, is assigned as the systematic uncertainty in the single-tag yield.

\subsubsection{Tracking and PID of $e^+$, $\mu^+$, and $\pi^+$}

The efficiencies for tracking and PID of the $e^+$ and $\mu^+$ are studied with the control samples of  $e^+e^-\to \gamma e^+e^-$ and  $e^+e^-\to \gamma \mu^+\mu^-$, respectively.
To consider the difference of topologies between $e^+e^-\to \gamma\ell^+\ell^-$ and $D^+_s\to \ell^+\nu_\ell$,
the obtained efficiencies in different polar angles and momentum intervals of the control samples are re-weighted to match the  $D^+_s\to \ell^+\nu_\ell$ signals.
The data-MC differences of the weighted efficiencies of $e^+$ tracking, and $e^+$ PID, $\mu^+$ tracking,  and  $\mu^+$ PID
are summarized in  Table~\ref{tab:trkPID}.
After correcting for the data-MC differences, the residual uncertainties are taken as individual systematic uncertainties, as listed in Table~\ref{tab:sys_crs}.

The efficiencies of the $\pi^+$ tracking and PID are studied with control samples of $e^+e^-\to K^+K^-\pi^+\pi^-(\pi^0)$ and $\pi^+\pi^-\pi^+\pi^-(\pi^0)$.
The systematic uncertainties of the $\pi^+$ tracking and PID efficiencies are assigned to be 0.35\% and 0.08\%, respectively.

\begin{table*}
\centering
\caption{Data-MC differences of the weighted efficiencies of tracking and PID for $e^+$ and $\mu^+$.\label{tab:trkPID}}
\begin{tabular}{lcccc}
\hline\hline
Signal decay                                                          &$e^+$ tracking&$e^+$ PID      &$\mu^+$ tracking&$\mu^+$ PID \\ \hline
$D^+_s\to \mu^{+}_a\nu_{\mu}$                                      & ... &...  & $99.93\pm0.12$ & $88.62\pm0.63$   \\
$D^+_s\to \mu^{+}_b\nu_{\mu}$                                      & ... &... & $99.93\pm0.12$ & $99.52\pm0.53$   \\
$D^+_s\to \tau^{+}_e\nu_{\tau}$     &$99.84\pm0.08$& $98.45\pm0.54$ &  ... &  ...  \\
$D^+_s\to \tau^{+}_\mu\nu_{\tau}$ & ...& ...  &  $99.93\pm0.12$ &  $89.05\pm1.06$    \\
 \hline\hline
\end{tabular}
\end{table*}

\subsubsection{$\gamma(\pi^0)$ selection}

The photon selection efficiency was previously studied with $J/\psi\to\pi^+\pi^-\pi^0$ decays~\cite{Upi0}.
The $\pi^0$ reconstruction efficiency was previously studied with $e^+e^-\to K^+K^-\pi^+\pi^-\pi^0$ events.
The systematic uncertainty of finding the transition $\gamma$ or $\pi^0$, weighted according to the BFs for $D_s^{*+}\to\gamma D_s^+$ and $D_s^{*+}\to\pi^0D_s^+$~\cite{pdg2024}, is 1.0\%. For the $\pi^0$ in the $D^+_s\to \tau^+_\rho \nu_\tau$, the systematic uncertainty is assigned to be 1.1\% from the study of Ref.~\cite{bes3_Ds_tauv1}. 
After re-weighting by the individual signal yields, the systematic uncertainty due to the $\pi^0$ in the $D^+_s\to \tau^+_\rho \nu_\tau$ for the overall BF measurement of  $D^+_s\to \tau^+\nu_\tau$ is 0.29\%.

\subsubsection{$E^{\mathrm{extra}~\gamma}_{\rm sum}$ and $N_{\rm extra}^{\rm charge}$ requirements}

The efficiency for the  requirements on $E^{\mathrm{extra}~\gamma}_{\rm sum}$ and
 $N_{\rm extra}^{\rm charge}$
is investigated with the double-tag sample of $D^+_s\to \eta \pi^+$.
The ratio of the averaged efficiency of data to that of simulation is $1.058\pm0.022$. After multiplying the signal efficiency by this factor, we assign 2.2\% as the systematic uncertainty.
Re-weighting by the individual signal yields, the systematic uncertainty due to $E^{\mathrm{extra}~\gamma}_{\rm sum}$ and $N_{\rm extra}^{\rm charge}$ requirements for the measurement of   $D^+_s\to \tau^+\nu_\tau$ is 0.56\%.

\subsubsection{Tag bias}
The single-tag efficiencies obtained from the inclusive MC sample differ from those
estimated with the signal MC events generated with events containing the single-tag $D_s^{*-}$ and
signal decays, an effect known as ``tag bias''.
To estimate the MC simulation for these differences, we use a method from Ref.~\cite{tagbias}.
To study the tag bias, we assign 1.0\% for the tracking and PID efficiencies of $\pi^+$ and $K^+$, 2.0\% for $\pi^0$, $K_S^0$, $\eta_{(\gamma\gamma)}$ reconstruction efficiencies of the tag side for their corresponding data and MC differences. The  difference from 1 of $\varepsilon_{\rm ST}^{D^+_s\to \mu_a^+\nu_\mu}/\varepsilon_{\rm ST}^{\rm inclusive~D_s^+}$ is assumed to not cancel in the BF measurements.
Weighting the offsets in each tag mode by their relative single-tag yield,  the average offset
for $D^+_s\to \mu_a^+\nu_\mu$ is calculated to be $(1.28\pm0.01)\%$, which is taken as the systematic uncertainty.

Similarly, the systematic uncertainties due to tag bias are assigned as 1.19\%, 1.28\%, 1.22\%, 1.19\%, and 0.68\% for $D^+_s\to \mu_b^+\nu_\mu$, $D^+_s\to \tau^+_e\nu_\tau$, $D^+_s\to \tau^+_\mu\nu_\tau$, $D^+_s\to \tau^+_\pi\nu_\tau$, and $D^+_s\to \tau^+_\rho\nu_\tau$, respectively. Re-weighting by individual signal yields, the systematic uncertainty due to tag bias for the measurement of the BF of $D^+_s\to \tau^+\nu_\tau$ is $1.16\%$.

\subsubsection{MC statistics}

The uncertainties due to the MC statistics are assigned to be 0.40\%, 0.51\%, 0.29\%, 0.40\%, and 0.32\% for $D^+_s\to\mu^+\nu_{\mu}$, $D^+_s\to \tau^+_e\nu_\tau$, $D^+_s\to \tau^+_\mu\nu_\tau$, $D^+_s\to \tau^+_\pi\nu_\tau$, and $D^+_s\to \tau^+_\rho\nu_\tau$, respectively. Re-weighting by individual signal yields, the systematic uncertainty due to MC statistics for the measurement of the BF of $D^+_s\to \tau^+\nu_\tau$ is $0.25\%$.

\subsubsection{$M^2_{\rm miss}$ and $E_{\rm sum}^{\rm extra~\gamma}$ fit}

The uncertainties in the $M^2_{\rm miss}$ and $E_{\rm sum}^{\rm extra~\gamma}$ fits arise from the signal  and background shapes.

For $D_s^+\to\mu^+_a\nu_\mu$, the systematic uncertainty due to the signal shape is estimated by replacing the
nominal shape with signal shape convolved with a double Gaussian function with floating parameters. The relative change between the re-measured and nominal BFs, 1.12\%, is taken as the systematic uncertainty.
For $\Dstotaunum$ and $\Dstotaunue$, we use the MC simulated shapes convolved with a single Gaussian resolution function with free parameters.
For $D_s\to\mu_b\nu_\mu$,  $\Dstotaunup$ and $\Dstotaunur$, the systematic uncertainty due to the signal shape is estimated by replacing the
nominal shape by varying the Gaussian shape parameters individually by $\pm 1\sigma$.
The quadrature sum of the relative changes between the re-measured BFs and the nominal BF, 0.60\%, is assigned as the systematic uncertainty for $D^+_s\to\tau^+\nu_\tau$.

For $D_s^+\to\mu^+_a\nu_\mu$, the peaking background is mainly due to the misidentification of a $\pi^+$ or a $K^+$ as a $\mu^+$.
We have corrected the background yields
considering the data-MC differences.
The systematic uncertainty is assigned by varying the weights of various background sources within $\pm1\sigma$ of individual BFs.
We also try alternative MC-simulated shapes by varying the relative fractions of the main backgrounds from $D_s^\pm D_s^{*\mp}$, $D_s^{*-}D_s^{*+}$ and $q\bar q$, by $\pm 1\sigma$ according to the observed cross sections Ref.~\cite{crsDsDss} and Ref.~\cite{crsDssDss}.
After considering the uncertainties of the correction factors and changing the weights of different background shapes according to their yield uncertainties, we assign 0.80\% as the associated systematic uncertainty.

To study the systematic uncertainty caused by the background shape in the fit for the measurement of $D^+_s\to\tau^+\nu_\tau$, we re-measure the BF by varying the background models in three ways. First, we use alternative MC-simulated shapes obtained by varying the relative fractions of the main backgrounds from $D_s^\pm D_s^{*\mp}$, $D_s^{*-}D_s^{*+}$ and $q\bar q$. We vary background yield by $\pm 1\sigma$ according to the observed cross sections~\cite{crsDsDss} and~\cite{crsDssDss}. Second,  we try alternative MC-simulated shapes obtained by varying the relative fractions of their largest background sources.
Third, we vary the yields of the main background sources by varying $\pm 1\sigma$ of the quoted BFs. Finally, we assign  2.34\% as the systematic uncertainty in the measurement of the BF of  $D^+_s\to\tau^+\nu_\tau$.

\subsubsection{Quoted branching fractions }

The BFs  of $\tmuvv$, $\tevv$, $\tpiv$ and $\trhov$ are quoted from the PDG~\cite{pdg2024}, which are $(17.39\pm0.04)\%$, $(17.82\pm0.04)\%$, $(10.82\pm0.05)\%$ and $(25.49\pm0.09)\%$, with relative uncertainties of 0.23\%, 0.22\%, 0.46\%, and 0.35\%, respectively. After re-weighting these by the individual signal yields, the systematic uncertainty for the BF of $D^+_s\to \tau^+\nu_\tau$ is assigned as $0.14\%$.

The BFs of $D_{s}^{*+}\to\gamma D^+_s$ and $D_{s}^{*+}\to \pi^0 D^+_s$ are ($93.5 \pm 0.7$)\% and ($5.8 \pm 0.7$)\% \cite{pdg2024}. By varying the BFs by these uncertainties, we find that the signal efficiencies change by no more than 0.30\%, which is assigned as a systematic uncertainty. The effect of the BF uncertainty on $D_{s}^{*+}\to e^+e^- D^+_s$ is negligible. 

The total systematic uncertainties for the quoted BFs of $D^+_s\to \mu^+\nu_\mu$ and $D^+_s\to \tau^+\nu_\tau$ are 0.30\% and 0.33\%, respectively.

\subsection{Total systematic uncertainties without  lepton universality constraint}

Table \ref{tab:sys_crs} summarizes all systematic uncertainties in the measurements of the BFs of $D_s^+\to\mu_a^+\nu_\mu$, $D_s^+\to\mu_b^+\nu_\mu$,   and $D^+_s\to\tau^+\nu_\tau$.
Assuming that all systematic uncertainties are independent with each other,
the total systematic uncertainties in the  measurements of the BFs  of $D_s^+\to\mu_a^+\nu_\mu$, $D_s^+\to\mu_b^+\nu_\mu$,  and $D^+_s\to\tau^+\nu_\tau$ are obtained
by summing in quadrature, giving 2.88\%, 4.13\%, and 3.59\%, respectively. 

\subsection{Total systematic uncertainty with lepton universality constraint}

When constraining the yields of $\muv$ and $D^+_s\to \tau^+\nu_\tau$, the uncertainties of $N_{\rm ST}$ and quoted BFs are fully correlated and all other uncertainties are independent. For the  independent system uncertainties,  we vary  the signal efficiency of the corresponding uncertainties $\pm1\sigma$ to obtain the new BF measurement. The systematic uncertainty due to lepton universality constraint
is assigned to be 0.04\% by varying the fixed $R$ within $\pm 1\sigma$. 
We assign the relative change of the BF as the corresponding uncertainty.
Table~\ref{tab:sys_crs} summarizes the systematic uncertainties for individual sources.
The total systematic uncertainty in the measurement of the BFs of $D^+_s\to\ell^+\nu_\ell$ (SM) is the quadrature sum which gives 3.74\%.

\begin{table*}[htbp]\centering
	\caption{Relative systematic uncertainties (\%) in the measurements of the BFs  of $D_s^+\to\mu^+\nu_\mu$ and $D^+_s\to\tau^+\nu_\tau$.}
	\label{tab:sys_crs}
	\begin{tabular}{ccccc}
		\hline\hline
		Source          &   $D_{s}^{+}\to \mu^{+}_a\nu_{\mu}$ & $D_{s}^{+}\to \mu^{+}_b\nu_{\mu}$ &     $D^{+}_{s} \to \tau^+\nu_{\tau}$   &   $D^+_s\to\ell^+\nu_\ell$ (SM)\\ \hline
		Single-tag yield    &              1.88                 &1.88    &     1.88                                                         &     1.88   \\
		$\mu^+$ tracking&              0.19                 &0.19    &     0.04                                                         &     0.08    \\
		$\mu^+$ PID     &              0.63                 &1.06    &     0.21                                                         &     0.99     \\
		$e^+$ tracking  &         ...                       & ...      &     0.11                                                         &     0.10      \\
		$e^+$ PID       &        ...                        &...    &     0.24                                                         &     0.18       \\
		$\pi^+$ tracking&         ...                        &...    &        0.35                                                      &      0.18       \\
		$\pi^+$ PID     &         ...                        &...    &       0.08                                                       &      0.04 \\
		$\gamma/\pi^0$ reconstruction&                     1.00      &1.00   &       1.00                                               &       1.00\\
		$\pi^0$ reconstruction from $\rho^+$&                  ...       &...        &       0.29                                     &       0.19\\
		$E^{\rm extra ~\gamma}_{\rm sum}$ and $N^{\rm charge}_{\rm extra}$ requirements&      ...    & 0.40&     0.56  &      0.37\\
		$M^2_{\rm miss}$ fit and $E^{\rm extra ~\gamma}_{\rm sum}$ fit&               1.20              &  3.03   &            2.45                               &       2.67                        \\
		Tag bias        &                 1.28                      &1.19&  1.16                                                  &             0.83                       \\
		MC statistics   &                 0.40                 &   0.40&          0.25                                                   &               0.22
		\\
		Quoted BFs       &                 0.30           &  0.46    &             0.33                                                  &             0.33         \\

		SM constraint   &                 ...                &   ...&          ...                                                 &               0.04\\
		\hline
		Total        &2.88&4.13 &3.59&3.74\\
		
		\hline\hline
	\end{tabular}
\end{table*}
\section{CONCLUSION}

By analyzing $10.64~\mathrm{fb}^{-1}$ of $e^+e^-$ collision data taken at $E_{\rm cm}$ between $4.237$ and $4.699$ GeV with the BESIII detector,
the BFs of  $D_s^+\to\mu^+\nu_\mu$ and $D_s^+\to\tau^+\nu_\tau$
are determined with and without constraining the ratios of their BFs to the SM prediction.
The obtained BFs are shown in Table~\ref{tab:branching fractions _data}.
Combining these BFs with the world averages of the masses of $\ell^+$ and $D_s^+$ as well as the lifetime of $D_s^+$,
we obtain $f_{D_s^+}|V_{cs}|$ with Eq.~\ref{eq01}. The comparison of the BFs and $f_{D_s^+}|V_{cs}|$ obtained 
in this work and the other measurements are shown in Table~\ref{tab:resultsummary}.

With the BFs of $D_s^+\to\mu^+\nu_\mu$ and $D_s^+\to\tau^+\nu_\tau$ without the SM constraint, we determine the ratio of the two
decay widths to be $\frac{\Gamma_{D_s^+\to\tau^+\nu_\tau}}{\Gamma_{D_s^+\to\mu^+\nu_\mu}} =10.24\pm0.57$. It is consistent with
the SM prediction of 9.75 within 0.9$\sigma$, implying no violation of $\tau-\mu$ lepton-flavor universality.
Taking the CKM matrix element $|V_{cs}|=0.97349\pm0.00016$ from the global SM fit
~\cite{pdg2024}, we obtain ${f_{D^+_s}}_{\mu\nu}=(\mufdsresult)$\,MeV and ${f_{D^+_s}}_{\tau\nu}=(\taufdsresult)$\,MeV, which agree with the value from recent LQCD calculations~\cite{prd98_074512}
within 0.4$\sigma$ and 1.6$\sigma$.  Conversely, taking the averaged decay constant
$f_{D_s^+}=(249.9 \pm 0.5)~\mathrm{MeV}$ from recent LQCD calculations~\cite{prd98_074512}, we obtain $|V_{cs}|_{\mu\nu} =\muvcsresult$ and
$|V_{cs}|_{\tau\nu} = \tauvcsresult$, which agree with the value from from the global 
SM fit ~\cite{pdg2024} within 0.4$\sigma$ and 1.6$\sigma$.
The $f_{D_s^+}$ values obtained in this work offer complementary data to test the LQCD calculations and the $|V_{cs}|$ measurements are important for CKM matrix unitarity tests.

The averaged  BF of $D_s^+\to\mu^+\nu_\mu$ is obtained by re-weighting the results reported in Refs.~\cite{bes4009,bes3_Ds_muv1} and the one in this work. 
The averaged BF of $D_s^+\to\tau^+\nu_\tau$ is obtained by re-weighting the results measured by using the decays  $\tau^+\to\pi^+\pi^0\bar{\nu}_\tau$~\cite{bes3_Ds_tauv1}, $\tau^+\to e^+\bar{\nu}_\tau\nu_e$~\cite{bes3_Ds_tauv3}, $\tau^+\to \mu^+\bar{\nu}_\tau\nu_\mu$~\cite{llcmvv},  $\tau^+\to\pi^+\bar{\nu}_\tau$~\cite{xiechenpiv}, the one measured in this work,  and  Ref.~\cite{bes4009}. Using the method described in~\cite{averagemethod} which takes into account the correlation of systematic uncertainties, we obtain the averaged BFs to be ${\mathcal B}_{D_s^+\to\mu^+\nu_\mu}= (0.5310\pm0.0099_{\rm stat}\pm0.0053_{\rm syst})\%$ and ${\mathcal B}_{D_s^+\to\tau^+\nu_\tau} = (5.359\pm0.067_{\rm stat}\pm0.074_{\rm syst})\%$.  
For $\muv$, the single-tag yield, the $\pi^+$ tracking and PID, the transition $\gamma(\pi^0)$ reconstruction are taken to be correlated; for $\tauv$, the uncertainties from the single-tag yield, the $\pi^+$ tracking and PID, the transition $\gamma(\pi^0)$ reconstruction, the best $\gamma(\pi^0)$ selection, and the tag bias are taken to be correlated. Additional common uncertainties come from $\tau_{D_s^+}$, $m_{D_s^+}$, and $m_{\tau,\mu}$ for $f_{D_s^+}$ and $|V_{cs}|$, while all the other uncertainties are independent. We obtain   ${f_{D_s^+}}_{\mu\nu} = (\Cmufdsresult)~\text{MeV}$ and ${f_{D_s^+}}_{\tau\nu} = (\Ctaufdsresult)~\text{MeV}$, which agree with the value from recent LQCD calculations~\cite{prd98_074512}
within 0.3$\sigma$ and 1.4$\sigma$. We also obtain $|V_{cs}|_{\mu\nu} = \Cmuvcsresult$,  and $|V_{cs}|_{\tau\nu} = \Ctauvcsresult$, which agree with the value from from the global SM fit~\cite{pdg2024} within 0.3$\sigma$ and 1.5$\sigma$. 

Furthermore, we re-weight the two separate $f_{D_s^+}|V_{cs}|$ values from $D^+_s\to \mu^+\nu_\mu$ and $D^+_s\to \tau^+\nu_\tau$  mentioned above, under the assumption that the uncertainties from the single-tag yield, the $\pi^+$ tracking and PID, the $e^+$ tracking and PID, the $\mu^+$ tracking and PID, the transition $\gamma(\pi^0)$ reconstruction, the best $\gamma(\pi^0)$ selection, and the tag bias are taken to be correlated. The common uncertainties come from $\tau_{D_s^+}$, $m_{D_s^+}$ and $m_{\tau,\mu}$ for $f_{D_s^+}$ and $|V_{cs}|$. Finally, we obtain   ${f_{D_s^+}} = (\Ctotfdsresult)~\text{MeV}$, which agrees with the value from recent LQCD calculations~\cite{prd98_074512} within 0.8$\sigma$,  and $|V_{cs}| = \Ctotvcsresult$, which agrees with the value from the global SM fit
~\cite{pdg2024} within 0.8$\sigma$.

\begin{table*}[htb]\centering
	\caption{Comparisons of the BFs  and the corresponding products of $f_{D_s^+}|V_{cs}|$ from various experiments. The ``weighted'' values are obtained by combining results after considering the correlated effects.  
		The ``average" values  are obtained by weighting both statistical and systematic uncertainties, but not the third uncertainty dominated by the uncertainty of the $D_s^+$ lifetime. The uncertainties of ``average'' BFs  and the first uncertainties of ``average"  $f_{D^+_s}|V_{cs}|$ are the total experimental uncertainties combined from statistical and systematic effects, and the second uncertainties of ``average"  $f_{D^+_s}|V_{cs}|$ is due to the input uncertainty of the quoted lifetime of $D^+_s$. All ``weighted'' and ``average'' results do not include results constrained by the SM.}
	\label{tab:resultsummary}
	\def\1#1#2{\multicolumn{#1}{#2}}
	\resizebox{1.0\textwidth}{!}{
		\begin{tabular}{m{3.6cm} m{2.0cm} m{1.8cm} m{3.7cm} m{3.8cm} m{3.5cm}}
			\hline\hline
			Experiment & $E_{\rm cm}$ (GeV) & Mode & $D_s^+$ decay & $\mathcal B$~(\%)& $f_{D_s^+}|V_{\rm cs}|$ (MeV)\\
			\hline
			{\bf This work}   &\bf 4.237-4.699& $\pmb{\dstpdstm}$  & $\pmb{\tau^+_e\nu_\tau,\tau^+_\mu\nu_\tau, \tau^+_\pi \nu_\tau,  \tau^+_\rho\nu_\tau}$  & $\pmb{5.60\pm0.16\pm0.20}$ & $\pmb{252.7\pm3.6\pm4.5\pm0.6}$ \\
			{\bf This work (SM)}   & \bf 4.237-4.699& $\pmb{\dstpdstm}$  & $\pmb{\tau^+_e\nu_\tau,\tau^+_\mu\nu_\tau, \tau^+_\pi \nu_\tau,  \tau^+_\rho\nu_\tau}$  & $\pmb{5.39\pm0.14\pm0.20}$ & $\pmb{247.9\pm3.2\pm4.6\pm0.5}$ \\	
			BESIII~\cite{bes4009} & 4.009 &  $D_s^+D_s^-$  & $\tau^+_\pi \nu_\tau$   & $3.28\pm1.83\pm0.37$ & $193.4\pm53.9\pm10.9\pm0.5$ \\
			BESIII~\cite{bes3_Ds_tauv1} & 4.178-4.226 &  $D^\pm_sD^{*\mp}_s$ & $\tau^+_\rho\nu_\tau$  & $5.30\pm0.25\pm0.20$ & $245.8\pm5.8\pm4.6\pm0.5$ \\
				BESIII~\cite{bes3_Ds_tauv2} & 4.178-4.226 &  $D^\pm_sD^{*\mp}_s$  & $\tau^+_\pi \nu_\tau$  & $5.21\pm0.25\pm0.17$ & $243.7\pm5.8\pm4.0\pm0.5$ \\
				BESIII~\cite{bes3_Ds_tauv3} & 4.178-4.226 & $D^\pm_sD^{*\mp}_s$ & $\tau^+_e\nu_\tau$  & $5.27\pm0.10\pm0.13$ & $245.1\pm2.3\pm3.0\pm0.5$ \\
				BESIII~\cite{xiechenpiv}    &  4.128-4.226 & $D^\pm_sD^{*\mp}_s$  & $\tau^+_\pi\bar{\nu}_\tau$  & $5.44\pm0.17\pm0.13$ & $249.0\pm3.9\pm3.0\pm0.5 $ \\
				BESIII~\cite{llcmvv} & 4.128-4.226 &  $D^\pm_sD^{*\mp}_s$ & $\tau^+_\mu\bar{\nu}_\tau$  & $5.37\pm0.17\pm0.15$ & $247.4\pm3.9\pm3.5\pm0.5$ \\
			\hline
		    Weighted$^{a}$   & {$\cdot\cdot\cdot$ }& { $\cdot\cdot\cdot$}  &{$\tau^+\nu_\tau$ }  & {$5.359\pm0.067\pm0.075$}  &{ $247.2\pm1.5\pm1.7\pm0.5$ }\\ 
			\hline
			CLEO~\cite{cleo2009b}   & 4.170 & $D^\pm_sD^{*\mp}_s$   & $\tau^+_e\nu_\tau$ & $5.30\pm0.47\pm0.22$ & $245.8\pm10.9\pm5.1\pm0.5$ \\
			CLEO~\cite{cleo2009a}  & 4.170 &  $D^\pm_sD^{*\mp}_s$  & $\tau^+_\rho\nu_\tau$ & $5.52\pm0.57\pm0.21$ & $250.9\pm13.0\pm4.8\pm0.6$ \\
			CLEO~\cite{cleo2009}  & 4.170 & $D^\pm_sD^{*\mp}_s$   & $\tau^+_\pi \nu_\tau$   & $6.42\pm0.81\pm0.18$ & $270.5\pm17.1\pm3.8\pm0.6$ \\
			BaBar~\cite{babar2010}  & 10.56 &  $DKX\gamma D^{-}_s$ & $\tau^+_e\nu_\tau ,\tau^+_\mu\nu_\tau$ & $4.96\pm0.37\pm0.57$ & $237.8\pm8.9\pm13.7\pm0.5$ \\
			Belle~\cite{belle2013}  & 10.56 &  $DKX\gamma D^{-}_s$   & $\tau^+_\pi \nu_\tau, \tau^+_e\nu_\tau,\tau^+_\mu\nu_\tau$ & $5.70\pm0.21^{+0.31}_{-0.30}$ & $254.9\pm4.7\pm7.0\pm0.6$\\
			\hline
			Average&&&&$5.38\pm0.09$&$247.7\pm2.1\pm0.5$\\
			\hline
			\hline
			{\bf This work}   & \bf 4.237-4.699&$\pmb \dstpdstm$    & $\pmb{\mu^+_a\nu_\mu}$ & $\pmb{0.547\pm0.026\pm0.016}$ & $\pmb{246.5\pm5.9\pm3.6\pm0.5}$  \\
			{\bf This work (SM)}   & \bf 4.237-4.699&$\pmb \dstpdstm$    & $\pmb{\mu^+\nu_\mu}$ & $\pmb{0.553\pm0.014\pm0.020}$ &   $\cdot\cdot\cdot$\\
			BESIII~\cite{bes4009}  & 4.009 &  $D_s^+D_s^-$ & $\mu^+\nu_\mu$ & $0.517\pm0.075\pm0.021$ & $239.6\pm17.4\pm4.9\pm0.5$ \\
			BESIII \cite{bes3_Ds_muv}  & 4.178 &  $D_s^{\pm}D_s^{*\mp}$ & $\mu^+\nu_\mu$ & $0.549\pm0.016\pm0.015$ & $246.9\pm3.6\pm3.4\pm0.5$ \\
			BESIII~\cite{bes3_Ds_tauv2}    &4.178-4.226 &$D_s^{\pm}D_s^{*\mp}$ & $\mu^+\nu_\mu$&$0.535\pm0.013\pm0.016$  &$243.7\pm3.0\pm3.6\pm0.5$\\
			BESIII~\cite{bes3_Ds_muv1}  & 4.128-4.226&  $D_s^{\pm}D_s^{*\mp}$ & $\mu^+\nu_\mu$ & $0.5294\pm0.0108\pm0.0085$ & $242.5\pm2.5\pm1.9\pm0.5$ \\
			\hline
			Weighted$^{b}$   & {$\cdot\cdot\cdot$ }& { $\cdot\cdot\cdot$  }&{ $\mu^+\nu_\mu$  }& {$0.5310\pm0.0099\pm0.0053$ } & {$242.8\pm2.3\pm1.2\pm0.5$ }\\
			Weighted$^{c}$  & {$\cdot\cdot\cdot$} & { $\cdot\cdot\cdot$  }&   { $\tau^+\nu_\tau,\mu^+\nu_\mu$}& {$\cdot\cdot\cdot$}& {$245.4\pm1.3\pm1.7\pm0.5
				$ }\\ 
		
			\hline
			CLEO~\cite{cleo2009}   & 4.170 & $D^\pm_sD^{*\mp}_s$     & $\mu^+\nu_\mu$ & $0.565\pm0.045\pm0.017$ & $250.5\pm10.0\pm3.8\pm0.5$ \\
			BaBar~\cite{babar2010}   & 10.56 &  $DKX\gamma D^{-}_s$   & $\mu^+\nu_\mu$ & $0.602\pm0.038\pm0.034$ & $258.6\pm8.2\pm7.3\pm0.5$ \\
			Belle~\cite{belle2013}   & 10.56 &  $DKX\gamma D^{-}_s$     & $\mu^+\nu_\mu$ & $0.531\pm0.028\pm0.020$ & $242.8\pm6.4\pm4.6\pm0.5$ \\
			\hline	
			Average&&&&$0.539\pm0.009$&$244.6 \pm2.0\pm0.5$\\
			\hline\hline
		\end{tabular}
	}
	\begin{tablenotes}
		\footnotesize
	\item{Weighted$^a$ excludes ``BESIII~\cite{bes3_Ds_tauv2}''.}
	\item{Weighted$^b$ excludes ``BESIII~\cite{bes3_Ds_muv,bes3_Ds_tauv2}''.}
	\item{Weighted$^c$ excludes ``BESIII~\cite{bes3_Ds_tauv2,bes3_Ds_muv,bes3_Ds_tauv2}''.}
	\end{tablenotes}
\end{table*}

\section{Acknowledgement}
The BESIII Collaboration thanks the staff of BEPCII and the IHEP computing center for their strong support. This work is supported in part by National Key R\&D Program of China under Contracts Nos. 2023YFA1606000, 2023YFA1606704, 2020YFA0406300, 2020YFA0406400; National Natural Science Foundation of China (NSFC) under Contracts Nos. 12375092, 11635010, 11735014, 11935015, 11935016, 11935018, 11961141012, 12025502, 12035009, 12035013, 12061131003, 12192260, 12192261, 12192262, 12192263, 12192264, 12192265, 12221005, 12225509, 12235017; the Chinese Academy of Sciences (CAS) Large-Scale Scientific Facility Program; the CAS Center for Excellence in Particle Physics (CCEPP); Joint Large-Scale Scientific Facility Funds of the NSFC and CAS under Contract No. U1832207; 100 Talents Program of CAS; The Institute of Nuclear and Particle Physics (INPAC) and Shanghai Key Laboratory for Particle Physics and Cosmology; German Research Foundation DFG under Contracts Nos. 455635585, FOR5327, GRK 2149; Istituto Nazionale di Fisica Nucleare, Italy; Ministry of Development of Turkey under Contract No. DPT2006K-120470; National Research Foundation of Korea under Contract No. NRF-2022R1A2C1092335; National Science and Technology fund of Mongolia; National Science Research and Innovation Fund (NSRF) via the Program Management Unit for Human Resources \& Institutional Development, Research and Innovation of Thailand under Contract No. B16F640076; Polish National Science Centre under Contract No. 2019/35/O/ST2/02907; The Swedish Research Council; U. S. Department of Energy under Contract No. DE-FG02-05ER41374.

\section{Appendix A: Fits to the $M_{\rm BC}$ distributions of data single-tag yields in data, single-tag efficiencies, and double-tag efficiencies  at other energy points}

\begin{figure*}[htbp]
	\begin{center}
		\includegraphics[width=0.8\textwidth] {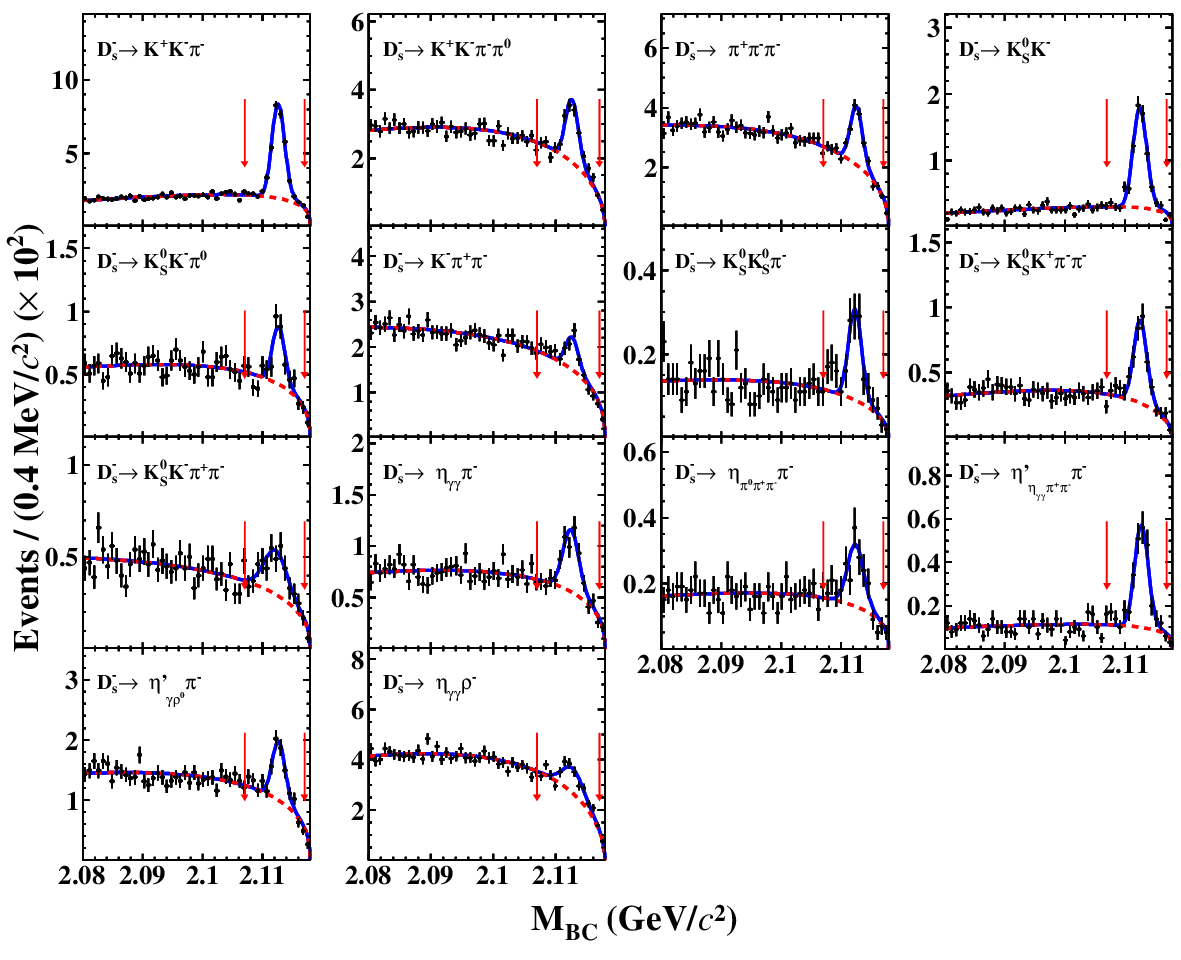}
		\caption{
			Fits to the $M_{\rm BC}$ distributions of ST $D^{*-}_s$ candidates selected from data at 4.237 GeV, where the points with error bars are data, the solid curves show the best fits, and the red dashed curves show the combinatorial background shapes. The pairs of arrows denote the $M_{\rm BC}$ signal window.}
					\label{fig:ST1}
	\end{center}
\end{figure*}

\begin{figure*}[htbp]
	\begin{center}\label{fig:ST2}
		\includegraphics[width=0.8\textwidth] {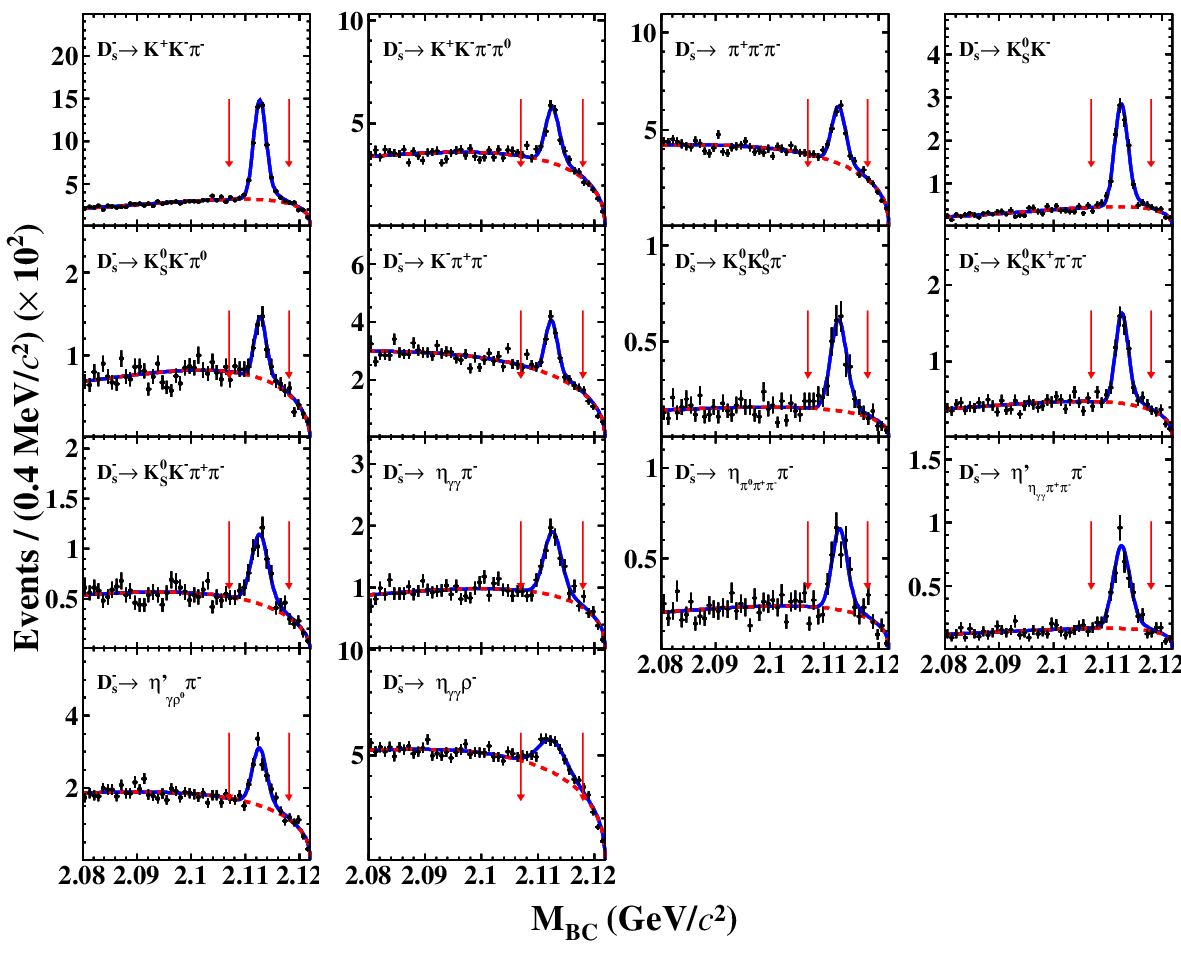}
		\caption{
			Fits to the $M_{\rm BC}$ distributions of ST $D^{*-}_s$ candidates selected from data at 4.246 GeV, where the points with error bars are data, the solid curves show the best fits, and the red dashed curves show the combinatorial background shapes. The pairs of arrows denote the $M_{\rm BC}$ signal window.}
	\end{center}
\end{figure*}

\begin{figure*}[htbp]
	\begin{center}\label{fig:ST3}
		\includegraphics[width=0.8\textwidth] {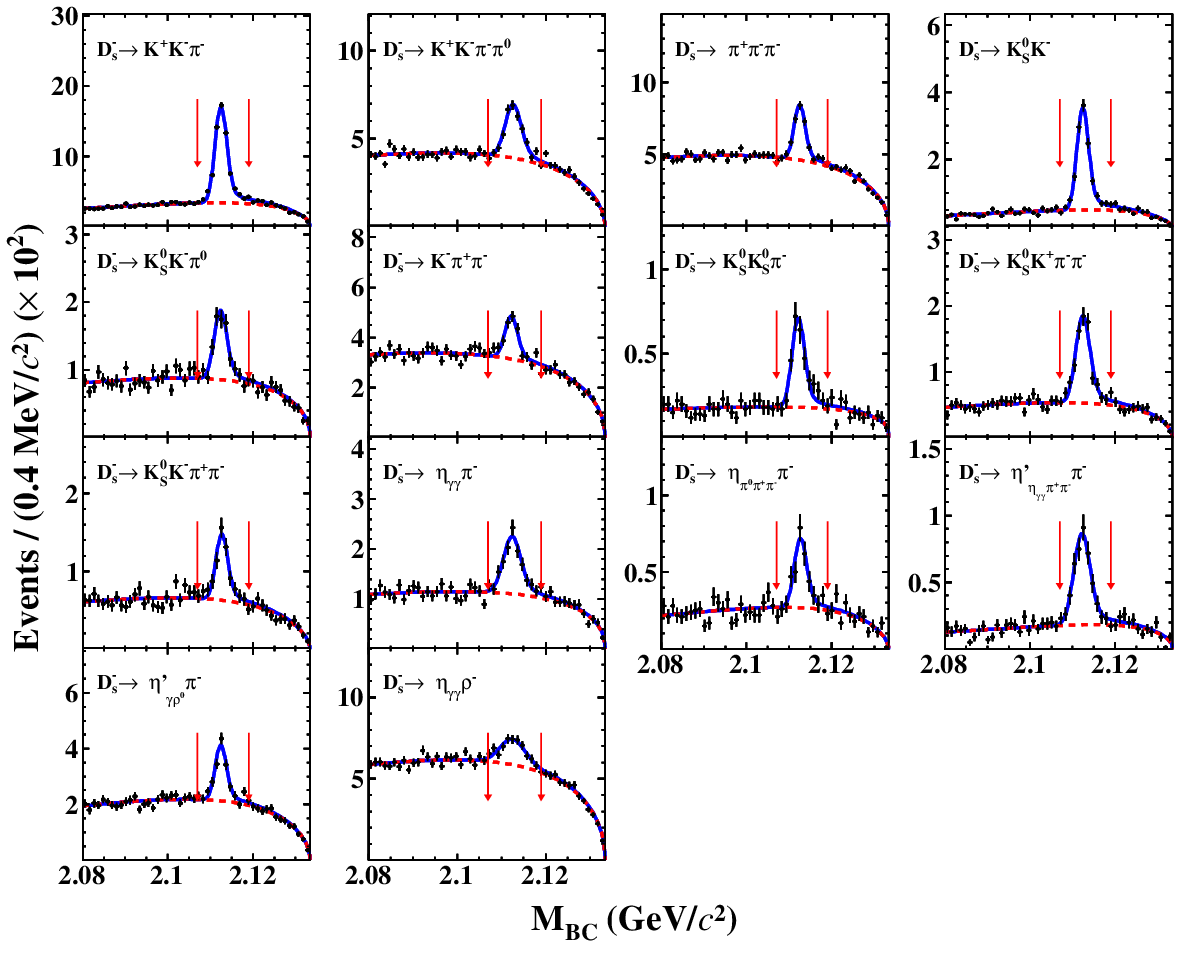}
		\caption{
			Fits to the $M_{\rm BC}$ distributions of ST $D^{*-}_s$ candidates selected from data at 4.270 GeV, where the points with error bars are data, the solid curves show the best fits, and the red dashed curves show the combinatorial background shapes. The pairs of arrows denote the $M_{\rm BC}$ signal window.}
	\end{center}
\end{figure*}

\begin{figure*}[htbp]
	\begin{center}\label{fig:ST4}
		\includegraphics[width=0.8\textwidth] {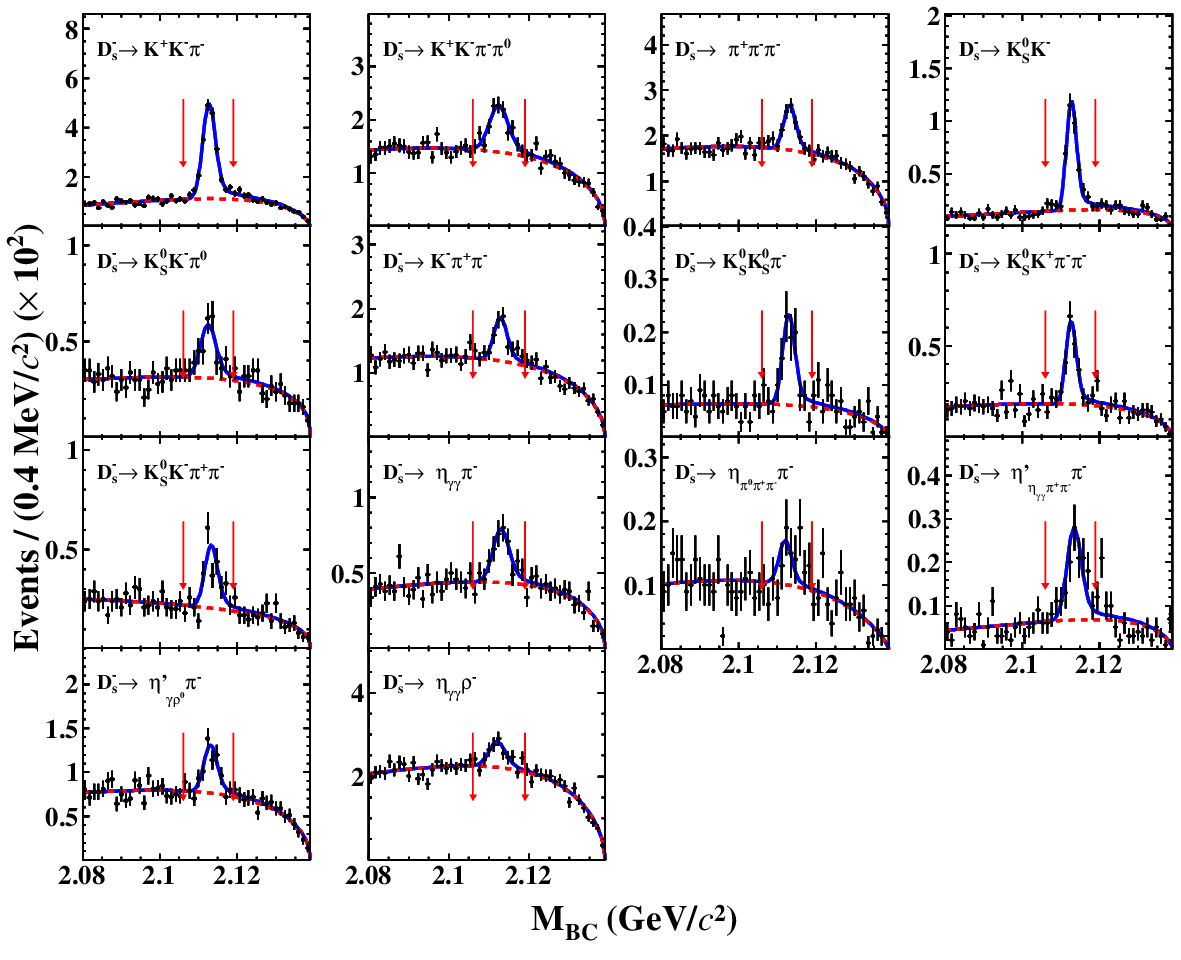}
		\caption{
			Fits to the $M_{\rm BC}$ distributions of ST $D^{*-}_s$ candidates selected from data at 4.280 GeV, where the points with error bars are data, the solid curves show the best fits, and the red dashed curves show  the combinatorial background shapes. The pairs of arrows denote the $M_{\rm BC}$ signal window.}
	\end{center}
\end{figure*}

\begin{figure*}[htbp]
	\begin{center}\label{fig:ST5}
		\includegraphics[width=0.8\textwidth] {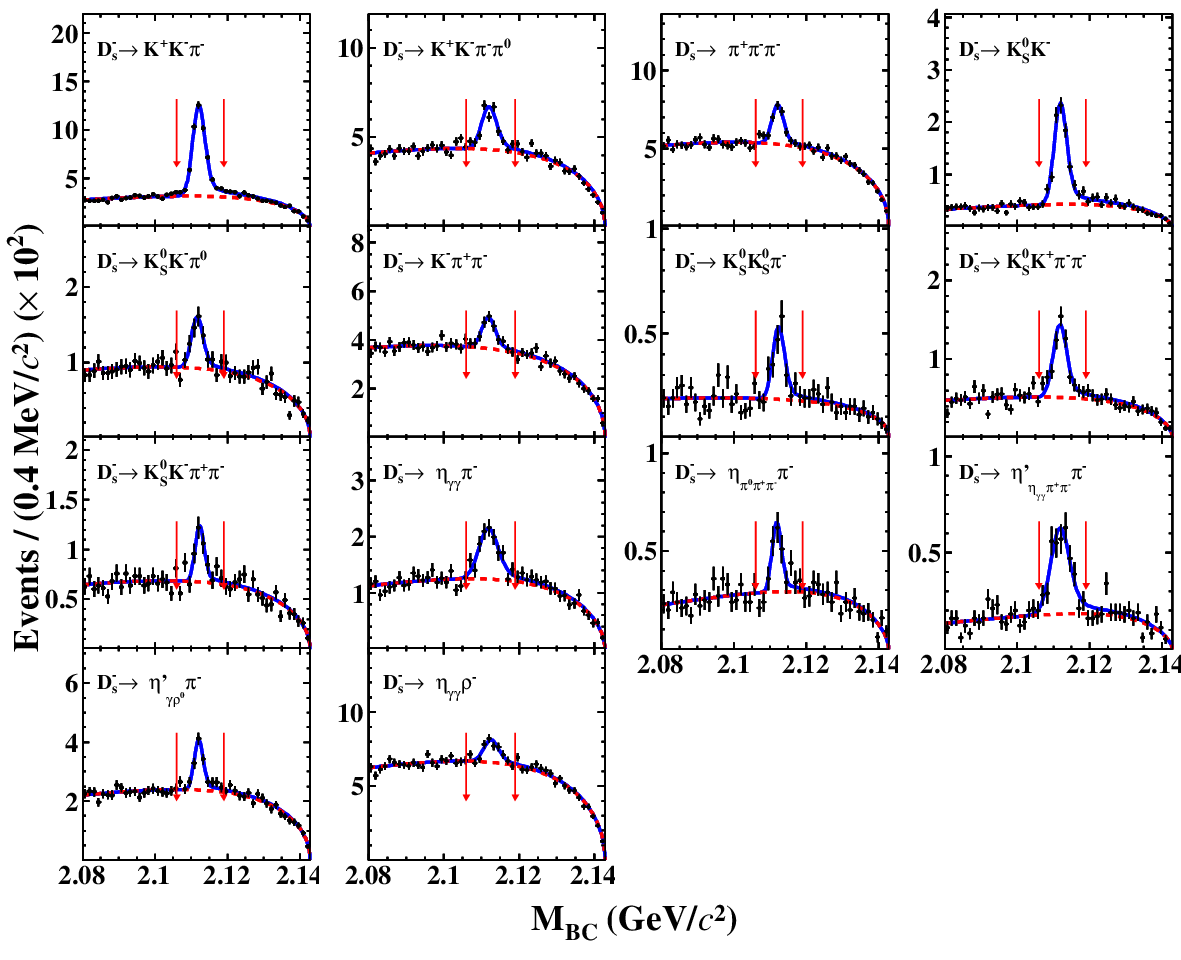}
		\caption{
			Fits to the $M_{\rm BC}$ distributions of ST $D^{*-}_s$ candidates selected from data at 4.290 GeV, where the points with error bars are data, the solid curves show the best fits, and the red dashed curves show  the combinatorial background shapes. The pairs of arrows denote the $M_{\rm BC}$ signal window.}
	\end{center}
\end{figure*}

\begin{figure*}[htbp]
	\begin{center}\label{fig:ST6}
		\includegraphics[width=0.8\textwidth] {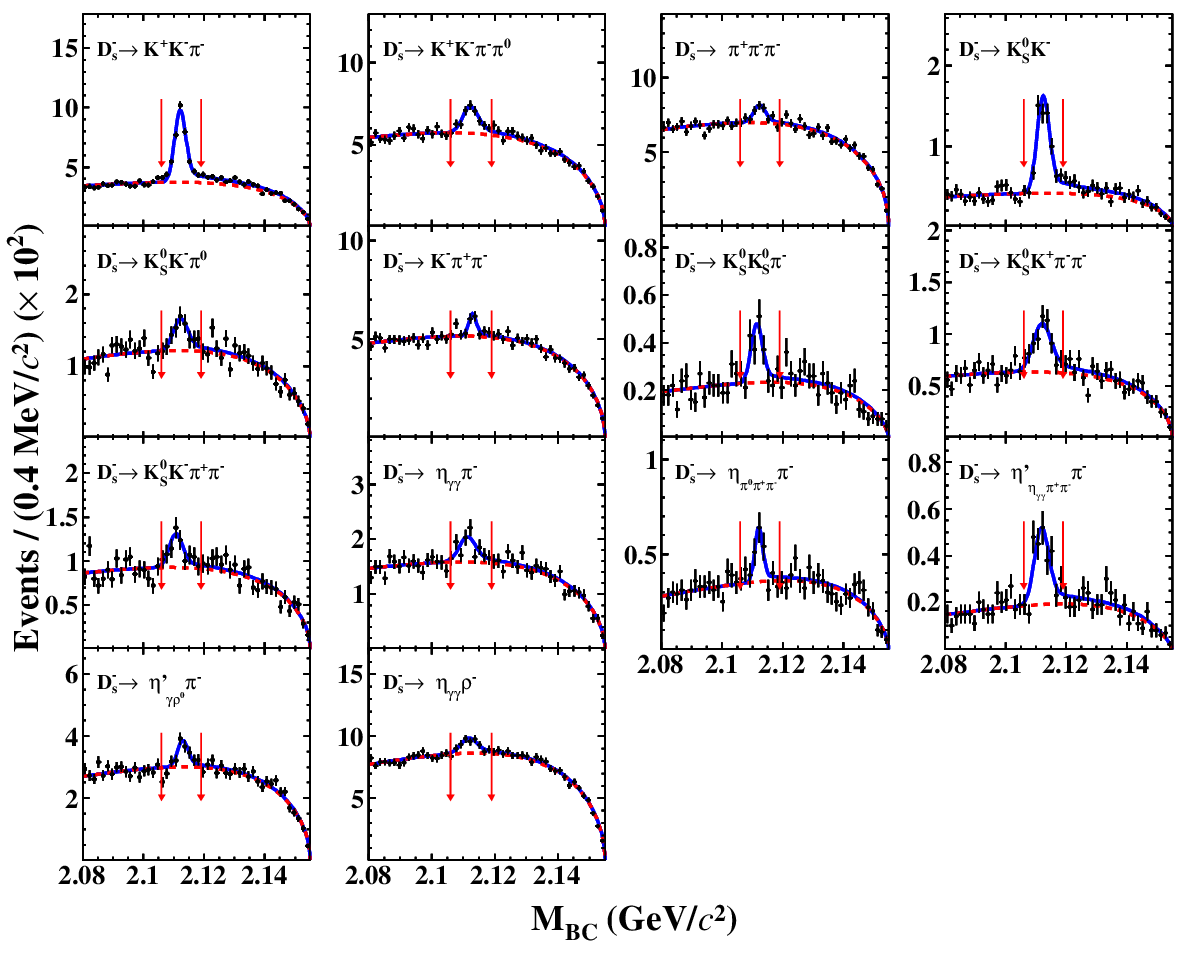}
		\caption{
			Fits to the $M_{\rm BC}$ distributions of ST $D^{*-}_s$ candidates selected from data at 4.310-4.315 GeV, where the points with error bars are data, the solid curves show the best fits, and the red dashed curves show  the combinatorial background shapes. The pairs of arrows denote the $M_{\rm BC}$ signal window.}
	\end{center}
\end{figure*}

\begin{figure*}[htbp]
	\begin{center}\label{fig:ST7}
		\includegraphics[width=0.8\textwidth] {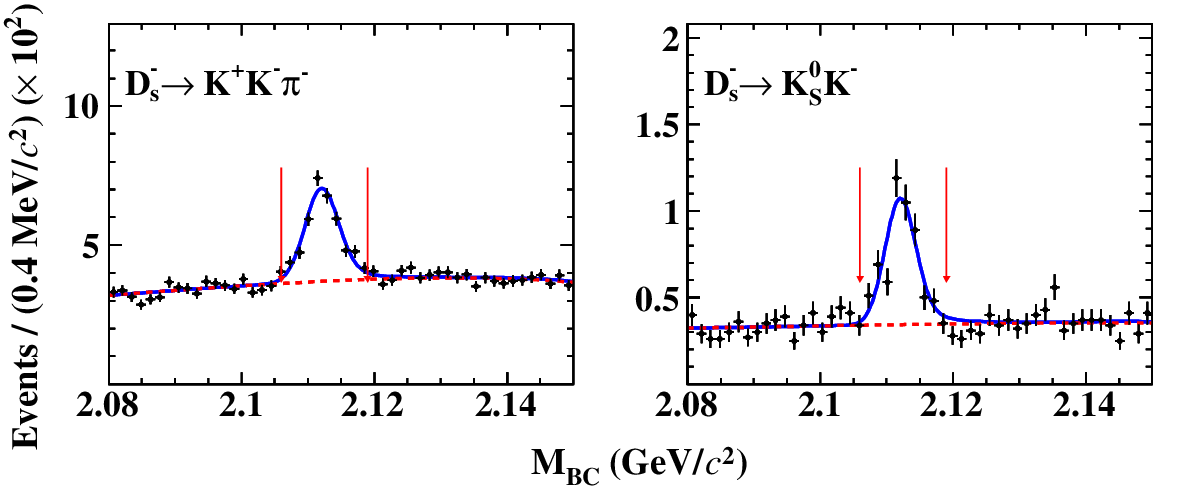}
		\caption{
			Fits to the $M_{\rm BC}$ distributions of ST $D^{*-}_s$ candidates selected from data at 4.400 GeV, where the points with error bars are data, the solid curves show the best fits, and the red dashed curves show the combinatorial background shapes. The pairs of arrows denote the $M_{\rm BC}$ signal window.}
	\end{center}
\end{figure*}

\begin{figure*}[htbp]
	\begin{center}\label{fig:ST8}
		\includegraphics[width=0.8\textwidth] {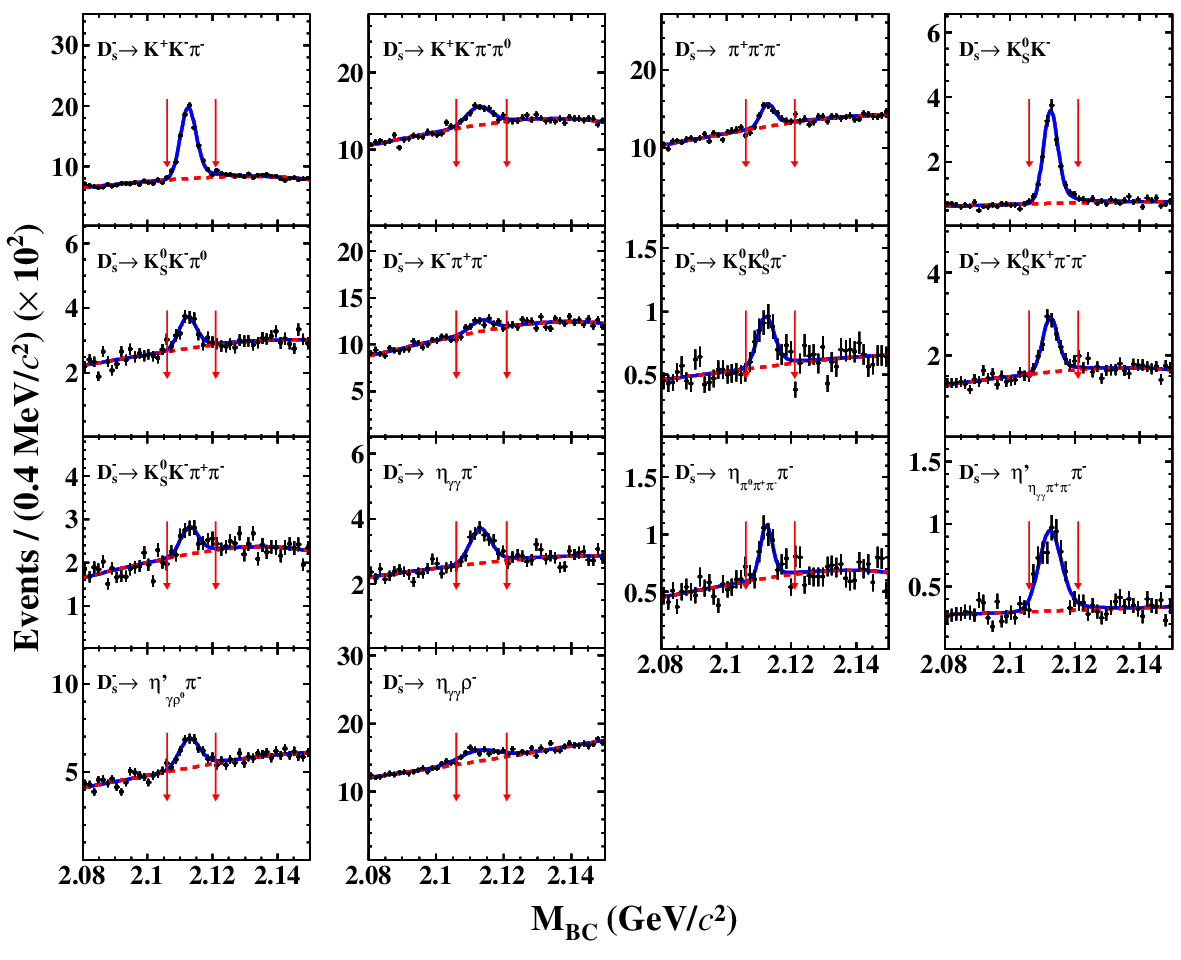}
		\caption{
			Fits to the $M_{\rm BC}$ distributions of ST $D^{*-}_s$ candidates selected from data at 4.420 GeV, where the points with error bars are data, the solid curves show the best fits, and the red dashed curves show the combinatorial background shapes. The pairs of arrows denote the $M_{\rm BC}$ signal window.}
	\end{center}
\end{figure*}

\begin{figure*}[htbp]
	\begin{center}\label{fig:ST9}
		\includegraphics[width=0.8\textwidth] {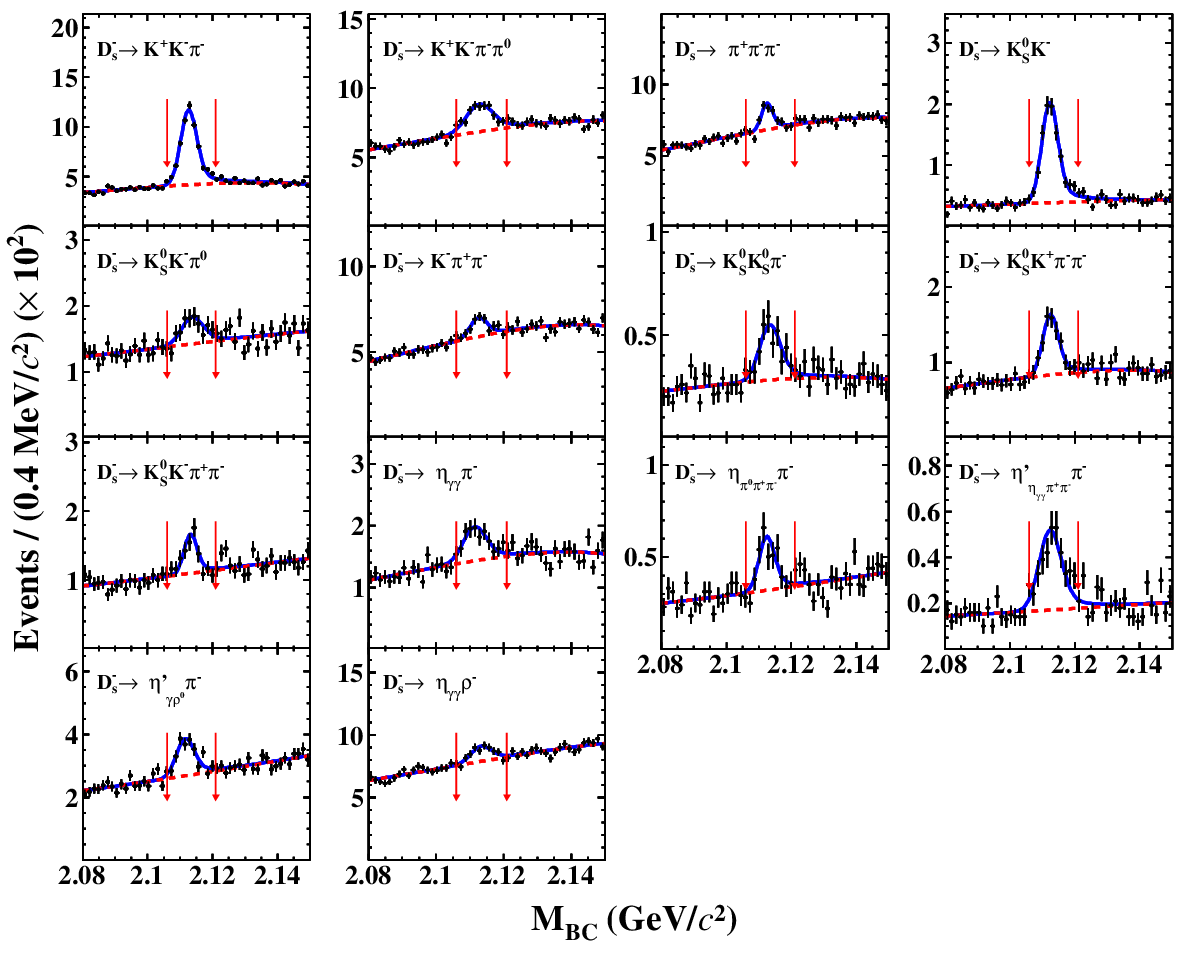}
		\caption{
			Fits to the $M_{\rm BC}$ distributions of ST $D^{*-}_s$ candidates selected from data at 4.440 GeV, where the points with error bars are data, the solid curves show the best fits, and the red dashed curves show the combinatorial background shapes. The pairs of arrows denote the $M_{\rm BC}$ signal window.}
	\end{center}
\end{figure*}

\begin{figure*}[htbp]
	\begin{center}
		\includegraphics[width=0.8\textwidth] {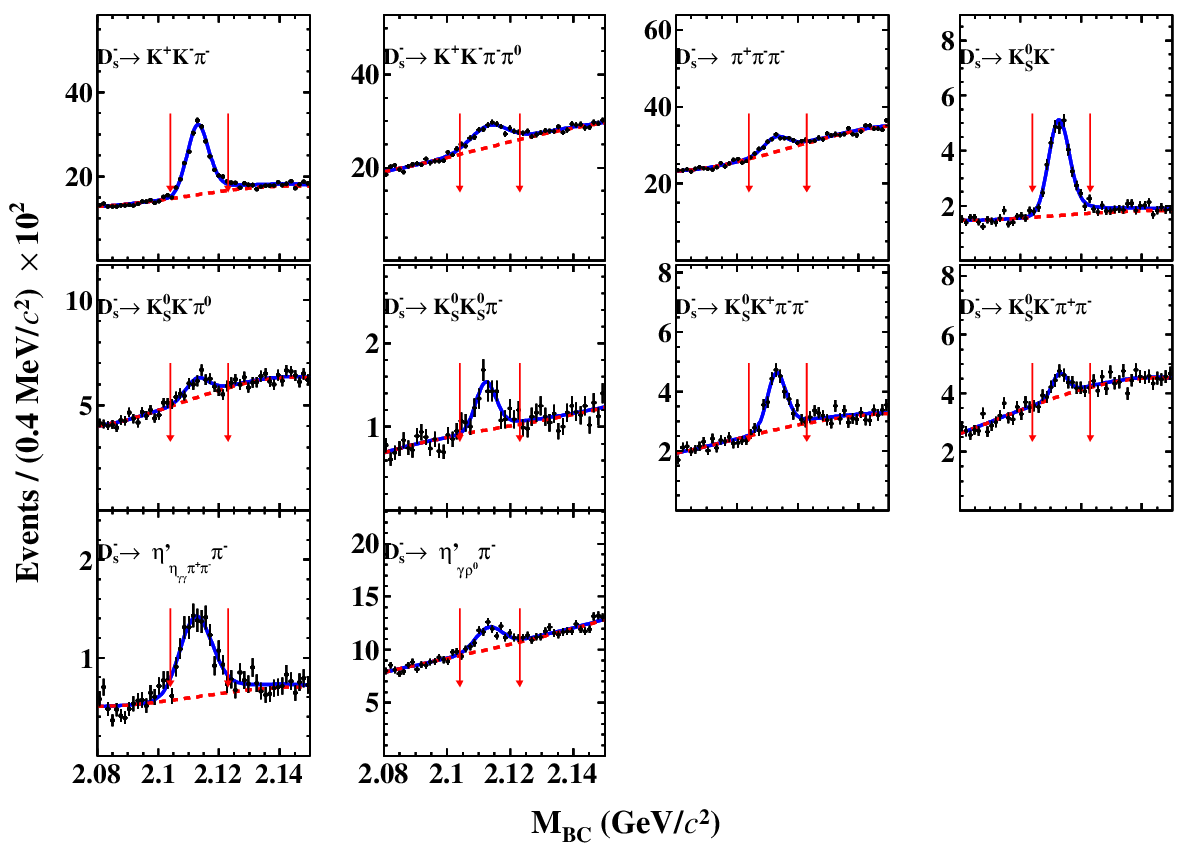}
		\caption{
			Fits to the $M_{\rm BC}$ distributions of ST $D^{*-}_s$ candidates selected from data at 4.470-4.699 GeV, where the points with error bars are data, the solid curves show the best fits, and the red dashed curves show the combinatorial background shapes. The pairs of arrows denote the $M_{\rm BC}$ signal window.}
			\label{fig:ST10}
	\end{center}
\end{figure*}

Figures \ref{fig:ST1}-\ref{fig:ST10} show the fits to the $M_{\rm BC}$ distributions of ST $D^{*-}_s$ candidates selected from data at the other energy points. Tables ~\ref{tab:ST1}-\ref{tab:ST10}
show the single-tag yields in data, single-tag efficiencies, and double-tag efficiencies at the other energy points. 

\begin{table*}[htbp]
	\caption{The single-tag yields ($N_{\rm ST}$), single-tag efficiencies ($\epsilon_{\rm ST}$), and double-tag efficiencies for each tag mode at 4.237~GeV. The $\epsilon^{\mu_a\nu}_{\rm DT}$ and $\epsilon^{\mu_b\nu}_{\rm DT}$ correspond to the double-tag efficiencies
		for $D^+_s\to \mu^+_a\nu_\mu$ and $D^+_s\to \mu^+_b\nu_\mu$; the $\epsilon^{\tau_e\nu}_{\rm DT}$, $\epsilon^{\tau_\mu\nu}_{\rm DT}$, $\epsilon^{\tau_\pi\nu}_{\rm DT}$, and $\epsilon^{\tau_\rho\nu}_{\rm DT}$ correspond to the double-tag efficiencies for
		$D^{+}_{s} \to \tau^+_e\nu_{\tau}$,
		$D^{+}_{s} \to \tau^+_\mu\nu_{\tau}$,
		$D^{+}_{s} \to \tau^+_\pi\nu_{\tau}$, and
		$D^{+}_{s} \to \tau^+_\rho\nu_{\tau}$, respectively. The efficiencies include the
		BFs of $D_s^{*+}$ and $\tau^+$ decays. The uncertainties are statistical only.
		\label{tab:ST1}}
	\centering
		\begin{tabular}{ l r@{} lr@{}  lr@{} lr@{} lr@{} lr@{} lr@{} lr@{} l}
			\hline\hline
			\makecell{$D^-_s$ tag \\mode} &\multicolumn{2}{c}{$N_{\rm ST}$} &\multicolumn{2}{c}{\makecell{$\epsilon_{\rm ST}$\\(\%)}}
			&\multicolumn{2}{c}{\makecell{$\epsilon^{\mu_a\nu}_{\rm DT}$\\(\%)}}&\multicolumn{2}{c}{\makecell{$\epsilon^{\mu_b\nu}_{\rm DT}$\\(\%)}}&\multicolumn{2}{c}{\makecell{$\epsilon^{\tau_e\nu}_{\rm DT}$\\(\%)}}&\multicolumn{2}{c}{\makecell{$\epsilon^{\tau_\mu\nu}_{\rm DT}$\\(\%)}}&\multicolumn{2}{c}{\makecell{$\epsilon^{\tau_\pi\nu}_{\rm DT}$\\(\%)}}&\multicolumn{2}{c}{\makecell{$\epsilon^{\tau_\rho\nu}_{\rm DT}$\\(\%)}} \\
			\hline
			$K^{+} K^{-}\pi^{-}$                              &$2303    $&$\pm72$&$20.47   $&$\pm0.13$&$16.46   $&$\pm0.08$&$17.98   $&$\pm0.09$&$3.92    $&$\pm0.02$&$1.83    $&$\pm0.01$&$1.76    $&$\pm0.01$&$1.1
			7    $&$\pm0.01$\\$K^{+} K^{-}\pi^{-}\pi^{0}$                       &$635     $&$\pm45$&$4.98    $&$\pm0.10$&$5.57    $&$\pm0.06$&$5.82    $&$\pm0.07$&$1.37    $&$\pm0.01$&$0.65    $&$\pm0.01$&$0.58    $&$\pm0.01$&$0.3
			4    $&$\pm0.01$\\$\pi^{-}\pi^{+}\pi^{-}$                           &$675     $&$\pm48$&$24.52   $&$\pm0.48$&$20.53   $&$\pm0.09$&$25.03   $&$\pm0.10$&$4.71    $&$\pm0.02$&$2.19    $&$\pm0.01$&$2.49    $&$\pm0.01$&$1.9
			4    $&$\pm0.02$\\$K_S^{0} K^{-}$                                   &$575     $&$\pm34$&$24.04   $&$\pm0.29$&$18.61   $&$\pm0.10$&$22.51   $&$\pm0.11$&$4.27    $&$\pm0.02$&$1.97    $&$\pm0.01$&$2.22    $&$\pm0.01$&$1.7
			1    $&$\pm0.02$\\$K_S^{0} K^{-}\pi^{0}$                            &$158     $&$\pm25$&$8.58    $&$\pm0.28$&$7.50    $&$\pm0.10$&$9.16    $&$\pm0.11$&$1.75    $&$\pm0.02$&$0.84    $&$\pm0.01$&$0.93    $&$\pm0.01$&$0.6
			1    $&$\pm0.02$\\$K^{-}\pi^{-}\pi^{+}$                             &$247     $&$\pm39$&$21.67   $&$\pm0.65$&$18.38   $&$\pm0.18$&$22.03   $&$\pm0.20$&$4.25    $&$\pm0.04$&$1.98    $&$\pm0.03$&$2.19    $&$\pm0.02$&$1.6
			0    $&$\pm0.03$\\$K_S^{0} K_S^{0}\pi^{-}$                          &$80      $&$\pm21$&$11.44   $&$\pm0.43$&$9.27    $&$\pm0.16$&$11.47   $&$\pm0.18$&$2.11    $&$\pm0.03$&$0.96    $&$\pm0.02$&$1.15    $&$\pm0.02$&$0.7
			2    $&$\pm0.02$\\$K_S^{0} K^{+}\pi^{-}\pi^{-}$                     &$235     $&$\pm28$&$10.51   $&$\pm0.20$&$8.36    $&$\pm0.13$&$9.89    $&$\pm0.14$&$1.99    $&$\pm0.03$&$0.93    $&$\pm0.02$&$0.94    $&$\pm0.01$&$0.6
			0    $&$\pm0.02$\\$K_S^{0} K^{-}\pi^{+}\pi^{-}$                     &$151     $&$\pm39$&$9.89    $&$\pm0.35$&$7.50    $&$\pm0.07$&$8.60    $&$\pm0.08$&$1.79    $&$\pm0.01$&$0.84    $&$\pm0.01$&$0.86    $&$\pm0.01$&$0.5
			4    $&$\pm0.01$\\$\eta_{\gamma\gamma}\pi^{-}$                      &$257     $&$\pm33$&$22.14   $&$\pm0.59$&$19.05   $&$\pm0.09$&$24.56   $&$\pm0.10$&$4.27    $&$\pm0.02$&$1.95    $&$\pm0.01$&$2.46    $&$\pm0.01$&$1.9
			9    $&$\pm0.02$\\$\eta_{\pi^{+}\pi^{-}\pi^{0}}\pi^{-}$             &$84      $&$\pm19$&$10.96   $&$\pm0.51$&$9.91    $&$\pm0.08$&$12.48   $&$\pm0.09$&$2.27    $&$\pm0.01$&$1.06    $&$\pm0.01$&$1.24    $&$\pm0.01$&$0.8
			9    $&$\pm0.01$\\$\eta\prime_{\pi^{+}\pi^{-}\eta} \pi^{-}$         &$179     $&$\pm20$&$11.79   $&$\pm0.32$&$10.06   $&$\pm0.07$&$12.94   $&$\pm0.08$&$2.27    $&$\pm0.01$&$1.06    $&$\pm0.01$&$1.29    $&$\pm0.01$&$0.9
			5    $&$\pm0.01$\\$\eta\prime_{\gamma\rho^{0}} \pi^{-}$             &$350     $&$\pm47$&$15.44   $&$\pm0.39$&$12.98   $&$\pm0.07$&$16.46   $&$\pm0.08$&$2.96    $&$\pm0.01$&$1.39    $&$\pm0.01$&$1.63    $&$\pm0.01$&$1.1
			2    $&$\pm0.01$\\$\eta_{\gamma\gamma}\rho^-$    &$548     $&$\pm83$&$9.07    $&$\pm0.27$&$9.31    $&$\pm0.06$&$12.59   $&$\pm0.07$&$2.12    $&$\pm0.01$&$1.01    $&$\pm0.01$&$1.29    $&$\pm0.01$&$0.9
			2    $&$\pm0.01$\\
			\hline\hline
		\end{tabular}
\end{table*}
\begin{table*}[htbp]
	\caption{The  single-tag yields ($N_{\rm ST}$), single-tag efficiencies ($\epsilon_{\rm ST}$), and double-tag efficiencies for each tag mode at 4.246~GeV. The $\epsilon^{\mu_a\nu}_{\rm DT}$ and $\epsilon^{\mu_b\nu}_{\rm DT}$ correspond to the double-tag efficiencies
		for $D^+_s\to \mu^+_a\nu_\mu$ and $D^+_s\to \mu^+_b\nu_\mu$; the $\epsilon^{\tau_e\nu}_{\rm DT}$, $\epsilon^{\tau_\mu\nu}_{\rm DT}$, $\epsilon^{\tau_\pi\nu}_{\rm DT}$, and $\epsilon^{\tau_\rho\nu}_{\rm DT}$ correspond to the double-tag efficiencies for
		$D^{+}_{s} \to \tau^+_e\nu_{\tau}$,
		$D^{+}_{s} \to \tau^+_\mu\nu_{\tau}$,
		$D^{+}_{s} \to \tau^+_\pi\nu_{\tau}$, and
		$D^{+}_{s} \to \tau^+_\rho\nu_{\tau}$, respectively. The efficiencies include the
		BFs of $D_s^{*+}$ and $\tau^+$ decays. The uncertainties are statistical only.
		\label{tab:ST2}}
	\centering
		\begin{tabular}{ l r@{} lr@{}  lr@{} lr@{} lr@{} lr@{} lr@{} lr@{} l }
			\hline\hline
			\makecell{$D^-_s$ tag \\mode}  &\multicolumn{2}{c}{$N_{\rm ST}$} &\multicolumn{2}{c}{\makecell{$\epsilon_{\rm ST}$\\(\%)}}
			&\multicolumn{2}{c}{\makecell{$\epsilon^{\mu_a\nu}_{\rm DT}$\\(\%)}}&\multicolumn{2}{c}{\makecell{$\epsilon^{\mu_b\nu}_{\rm DT}$\\(\%)}}&\multicolumn{2}{c}{\makecell{$\epsilon^{\tau_e\nu}_{\rm DT}$\\(\%)}}&\multicolumn{2}{c}{\makecell{$\epsilon^{\tau_\mu\nu}_{\rm DT}$\\(\%)}}&\multicolumn{2}{c}{\makecell{$\epsilon^{\tau_\pi\nu}_{\rm DT}$\\(\%)}}&\multicolumn{2}{c}{\makecell{$\epsilon^{\tau_\rho\nu}_{\rm DT}$\\(\%)}} \\
			\hline
			$K^{+} K^{-}\pi^{-}$                              &$4241    $&$\pm94$&$20.11   $&$\pm0.09$&$16.92 
			$&$\pm0.08$&$17.49   $&$\pm0.09$&$3.83    $&$\pm0.02$&$1.78    $&$\pm0.01$&$1.71    $&$\pm0.01$&$1.14    $&$\pm0.01$\\$K^{+} K^{-}\pi^{-}\pi^{0}$                       &$1029    $&$\pm81$&$5.39    $&$\pm0.07$&$5.82  
			$&$\pm0.06$&$5.64    $&$\pm0.07$&$1.34    $&$\pm0.01$&$0.63    $&$\pm0.01$&$0.56    $&$\pm0.01$&$0.32    $&$\pm0.01$\\$\pi^{-}\pi^{+}\pi^{-}$                           &$1149    $&$\pm78$&$25.80   $&$\pm0.34$&$20.65 
			$&$\pm0.09$&$24.81   $&$\pm0.10$&$4.59    $&$\pm0.02$&$2.12    $&$\pm0.01$&$2.45    $&$\pm0.01$&$1.85    $&$\pm0.02$\\$K_S^{0} K^{-}$                                   &$920     $&$\pm41$&$23.66   $&$\pm0.21$&$18.74 
			$&$\pm0.10$&$22.02   $&$\pm0.11$&$4.21    $&$\pm0.02$&$1.95    $&$\pm0.01$&$2.19    $&$\pm0.01$&$1.67    $&$\pm0.02$\\$K_S^{0} K^{-}\pi^{0}$                            &$254     $&$\pm31$&$7.89    $&$\pm0.18$&$7.44  
			$&$\pm0.10$&$8.90    $&$\pm0.11$&$1.71    $&$\pm0.02$&$0.78    $&$\pm0.01$&$0.89    $&$\pm0.01$&$0.60    $&$\pm0.02$\\$K^{-}\pi^{-}\pi^{+}$                             &$634     $&$\pm51$&$23.40   $&$\pm0.52$&$18.59 
			$&$\pm0.18$&$21.51   $&$\pm0.20$&$4.16    $&$\pm0.04$&$1.90    $&$\pm0.03$&$2.15    $&$\pm0.02$&$1.54    $&$\pm0.03$\\$K_S^{0} K_S^{0}\pi^{-}$                          &$216     $&$\pm23$&$11.21   $&$\pm0.28$&$9.26  
			$&$\pm0.16$&$11.10   $&$\pm0.18$&$2.05    $&$\pm0.03$&$0.96    $&$\pm0.02$&$1.09    $&$\pm0.02$&$0.76    $&$\pm0.03$\\$K_S^{0} K^{+}\pi^{-}\pi^{-}$                     &$433     $&$\pm33$&$10.08   $&$\pm0.15$&$8.53  
			$&$\pm0.13$&$9.36    $&$\pm0.14$&$2.00    $&$\pm0.03$&$0.90    $&$\pm0.02$&$0.92    $&$\pm0.01$&$0.59    $&$\pm0.02$\\$K_S^{0} K^{-}\pi^{+}\pi^{-}$                     &$330     $&$\pm35$&$9.07    $&$\pm0.23$&$7.65  
			$&$\pm0.07$&$8.44    $&$\pm0.07$&$1.77    $&$\pm0.01$&$0.84    $&$\pm0.01$&$0.83    $&$\pm0.01$&$0.53    $&$\pm0.01$\\$\eta_{\gamma\gamma}\pi^{-}$                      &$492     $&$\pm47$&$20.56   $&$\pm0.40$&$18.65 
			$&$\pm0.08$&$23.91   $&$\pm0.10$&$4.15    $&$\pm0.02$&$1.91    $&$\pm0.01$&$2.39    $&$\pm0.01$&$1.91    $&$\pm0.02$\\$\eta_{\pi^{+}\pi^{-}\pi^{0}}\pi^{-}$             &$192     $&$\pm22$&$10.92   $&$\pm0.34$&$9.90  
			$&$\pm0.07$&$12.42   $&$\pm0.07$&$2.20    $&$\pm0.01$&$1.02    $&$\pm0.01$&$1.22    $&$\pm0.01$&$0.86    $&$\pm0.01$\\$\eta\prime_{\pi^{+}\pi^{-}\eta} \pi^{-}$         &$311     $&$\pm27$&$11.35   $&$\pm0.22$&$10.03 
			$&$\pm0.07$&$12.82   $&$\pm0.07$&$2.20    $&$\pm0.01$&$1.03    $&$\pm0.01$&$1.25    $&$\pm0.01$&$0.91    $&$\pm0.01$\\$\eta\prime_{\gamma\rho^{0}} \pi^{-}$             &$650     $&$\pm56$&$14.82   $&$\pm0.25$&$12.70 
			$&$\pm0.07$&$15.83   $&$\pm0.08$&$2.89    $&$\pm0.01$&$1.34    $&$\pm0.01$&$1.56    $&$\pm0.01$&$1.08    $&$\pm0.01$\\$\eta_{\gamma\gamma}\rho^-$    &$1093    $&$\pm156$&$8.17    $&$\pm0.17$&$9.02 
			$&$\pm0.06$&$12.25   $&$\pm0.07$&$2.04    $&$\pm0.01$&$0.96    $&$\pm0.01$&$1.24    $&$\pm0.01$&$0.90    $&$\pm0.01$\\
			\hline\hline
		\end{tabular}
\end{table*}
\begin{table*}[htbp]
	\caption{The  single-tag yields ($N_{\rm ST}$), single-tag efficiencies ($\epsilon_{\rm ST}$), and double-tag efficiencies for each tag mode at 4.270~GeV. The $\epsilon^{\mu_a\nu}_{\rm DT}$ and $\epsilon^{\mu_b\nu}_{\rm DT}$ correspond to the double-tag efficiencies
		for $D^+_s\to \mu^+_a\nu_\mu$ and $D^+_s\to \mu^+_b\nu_\mu$; the $\epsilon^{\tau_e\nu}_{\rm DT}$, $\epsilon^{\tau_\mu\nu}_{\rm DT}$, $\epsilon^{\tau_\pi\nu}_{\rm DT}$, and $\epsilon^{\tau_\rho\nu}_{\rm DT}$ correspond to the double-tag efficiencies for
		$D^{+}_{s} \to \tau^+_e\nu_{\tau}$,
		$D^{+}_{s} \to \tau^+_\mu\nu_{\tau}$,
		$D^{+}_{s} \to \tau^+_\pi\nu_{\tau}$, and
		$D^{+}_{s} \to \tau^+_\rho\nu_{\tau}$, respectively. The efficiencies include the
		BFs of $D_s^{*+}$ and $\tau^+$ decays. The uncertainties are statistical only.
		\label{tab:ST3}}
	\centering
		\begin{tabular}{ l r@{} lr@{}  lr@{} lr@{} lr@{} lr@{} lr@{} lr@{}l }
			\hline\hline
			\makecell{$D^-_s$ tag \\mode} &\multicolumn{2}{c}{$N_{\rm ST}$} &\multicolumn{2}{c}{\makecell{$\epsilon_{\rm ST}$\\(\%)}}
			&\multicolumn{2}{c}{\makecell{$\epsilon^{\mu_a\nu}_{\rm DT}$\\(\%)}}&\multicolumn{2}{c}{\makecell{$\epsilon^{\mu_b\nu}_{\rm DT}$\\(\%)}}&\multicolumn{2}{c}{\makecell{$\epsilon^{\tau_e\nu}_{\rm DT}$\\(\%)}}&\multicolumn{2}{c}{\makecell{$\epsilon^{\tau_\mu\nu}_{\rm DT}$\\(\%)}}&\multicolumn{2}{c}{\makecell{$\epsilon^{\tau_\pi\nu}_{\rm DT}$\\(\%)}}&\multicolumn{2}{c}{\makecell{$\epsilon^{\tau_\rho\nu}_{\rm DT}$\\(\%)}} \\
			\hline
			$K^{+} K^{-}\pi^{-}$                              &$4822    $&$\pm101$&$19.53   $&$\pm0.09$&$15.48
			$&$\pm0.08$&$16.05   $&$\pm0.08$&$3.49    $&$\pm0.02$&$1.61    $&$\pm0.01$&$1.60    $&$\pm0.01$&$1.11    $&$\pm0.01$\\$K^{+} K^{-}\pi^{-}\pi^{0}$                       &$1368    $&$\pm92$&$4.70    $&$\pm0.06$&$5.25  
			$&$\pm0.06$&$5.28    $&$\pm0.06$&$1.18    $&$\pm0.01$&$0.57    $&$\pm0.01$&$0.51    $&$\pm0.01$&$0.31    $&$\pm0.01$\\$\pi^{-}\pi^{+}\pi^{-}$                           &$1265    $&$\pm81$&$25.40   $&$\pm0.33$&$19.13 
			$&$\pm0.09$&$22.82   $&$\pm0.10$&$4.20    $&$\pm0.02$&$1.95    $&$\pm0.01$&$2.26    $&$\pm0.01$&$1.76    $&$\pm0.02$\\$K_S^{0} K^{-}$                                   &$1055    $&$\pm44$&$23.09   $&$\pm0.20$&$17.40 
			$&$\pm0.10$&$20.55   $&$\pm0.11$&$3.90    $&$\pm0.02$&$1.77    $&$\pm0.01$&$2.02    $&$\pm0.01$&$1.59    $&$\pm0.02$\\$K_S^{0} K^{-}\pi^{0}$                            &$388     $&$\pm40$&$7.74    $&$\pm0.18$&$6.79  
			$&$\pm0.10$&$8.34    $&$\pm0.11$&$1.54    $&$\pm0.02$&$0.73    $&$\pm0.01$&$0.79    $&$\pm0.01$&$0.56    $&$\pm0.02$\\$K^{-}\pi^{-}\pi^{+}$                             &$618     $&$\pm72$&$23.46   $&$\pm0.56$&$17.21 
			$&$\pm0.18$&$19.86   $&$\pm0.19$&$3.81    $&$\pm0.03$&$1.77    $&$\pm0.02$&$1.94    $&$\pm0.02$&$1.46    $&$\pm0.03$\\$K_S^{0} K_S^{0}\pi^{-}$                          &$199     $&$\pm23$&$10.98   $&$\pm0.28$&$8.67  
			$&$\pm0.16$&$10.70   $&$\pm0.18$&$1.91    $&$\pm0.03$&$0.89    $&$\pm0.02$&$1.00    $&$\pm0.02$&$0.72    $&$\pm0.02$\\$K_S^{0} K^{+}\pi^{-}\pi^{-}$                     &$550     $&$\pm39$&$10.08   $&$\pm0.16$&$7.89  
			$&$\pm0.12$&$8.73    $&$\pm0.13$&$1.83    $&$\pm0.03$&$0.82    $&$\pm0.02$&$0.88    $&$\pm0.01$&$0.57    $&$\pm0.02$\\$K_S^{0} K^{-}\pi^{+}\pi^{-}$                     &$300     $&$\pm36$&$9.40    $&$\pm0.26$&$7.20  
			$&$\pm0.07$&$7.97    $&$\pm0.07$&$1.62    $&$\pm0.01$&$0.77    $&$\pm0.01$&$0.76    $&$\pm0.01$&$0.52    $&$\pm0.01$\\$\eta_{\gamma\gamma}\pi^{-}$                      &$557     $&$\pm54$&$19.53   $&$\pm0.41$&$16.79 
			$&$\pm0.08$&$21.41   $&$\pm0.09$&$3.72    $&$\pm0.02$&$1.72    $&$\pm0.01$&$2.13    $&$\pm0.01$&$1.74    $&$\pm0.01$\\$\eta_{\pi^{+}\pi^{-}\pi^{0}}\pi^{-}$             &$167     $&$\pm24$&$10.59   $&$\pm0.36$&$8.93  
			$&$\pm0.06$&$11.06   $&$\pm0.07$&$2.00    $&$\pm0.01$&$0.93    $&$\pm0.01$&$1.09    $&$\pm0.01$&$0.78    $&$\pm0.01$\\$\eta\prime_{\pi^{+}\pi^{-}\eta} \pi^{-}$         &$323     $&$\pm27$&$10.91   $&$\pm0.23$&$9.01  
			$&$\pm0.06$&$11.48   $&$\pm0.07$&$2.03    $&$\pm0.01$&$0.92    $&$\pm0.01$&$1.15    $&$\pm0.01$&$0.85    $&$\pm0.01$\\$\eta\prime_{\gamma\rho^{0}} \pi^{-}$             &$674     $&$\pm59$&$14.26   $&$\pm0.27$&$11.58 
			$&$\pm0.07$&$14.43   $&$\pm0.08$&$2.62    $&$\pm0.01$&$1.23    $&$\pm0.01$&$1.42    $&$\pm0.01$&$1.04    $&$\pm0.01$\\$\eta_{\gamma\gamma}\rho^-$    &$1033    $&$\pm120$&$8.19    $&$\pm0.18$&$7.93 
			$&$\pm0.06$&$10.71   $&$\pm0.07$&$1.79    $&$\pm0.01$&$0.84    $&$\pm0.01$&$1.11    $&$\pm0.01$&$0.79    $&$\pm0.01$\\
			\hline\hline
		\end{tabular}
\end{table*}
\begin{table*}[htbp]
	\caption{The  single-tag yields ($N_{\rm ST}$), single-tag efficiencies ($\epsilon_{\rm ST}$), and double-tag efficiencies for each tag mode at 4.280~GeV. The $\epsilon^{\mu_a\nu}_{\rm DT}$ and $\epsilon^{\mu_b\nu}_{\rm DT}$ correspond to the double-tag efficiencies
		for $D^+_s\to \mu^+_a\nu_\mu$ and $D^+_s\to \mu^+_b\nu_\mu$; the $\epsilon^{\tau_e\nu}_{\rm DT}$, $\epsilon^{\tau_\mu\nu}_{\rm DT}$, $\epsilon^{\tau_\pi\nu}_{\rm DT}$, and $\epsilon^{\tau_\rho\nu}_{\rm DT}$ correspond to the double-tag efficiencies for
		$D^{+}_{s} \to \tau^+_e\nu_{\tau}$,
		$D^{+}_{s} \to \tau^+_\mu\nu_{\tau}$,
		$D^{+}_{s} \to \tau^+_\pi\nu_{\tau}$, and
		$D^{+}_{s} \to \tau^+_\rho\nu_{\tau}$, respectively. The efficiencies include the
		BFs of $D_s^{*+}$ and $\tau^+$ decays. The uncertainties are statistical only.
		\label{tab:ST4}}
	\centering
		\begin{tabular}{ l r@{} lr@{}  lr@{} lr@{} lr@{} lr@{} lr@{} lr@{} l }
			\hline\hline
			\makecell{$D^-_s$ tag \\mode}  &\multicolumn{2}{c}{$N_{\rm ST}$} &\multicolumn{2}{c}{\makecell{$\epsilon_{\rm ST}$\\(\%)}}
			&\multicolumn{2}{c}{\makecell{$\epsilon^{\mu_a\nu}_{\rm DT}$\\(\%)}}&\multicolumn{2}{c}{\makecell{$\epsilon^{\mu_b\nu}_{\rm DT}$\\(\%)}}&\multicolumn{2}{c}{\makecell{$\epsilon^{\tau_e\nu}_{\rm DT}$\\(\%)}}&\multicolumn{2}{c}{\makecell{$\epsilon^{\tau_\mu\nu}_{\rm DT}$\\(\%)}}&\multicolumn{2}{c}{\makecell{$\epsilon^{\tau_\pi\nu}_{\rm DT}$\\(\%)}}&\multicolumn{2}{c}{\makecell{$\epsilon^{\tau_\rho\nu}_{\rm DT}$\\(\%)}} \\
			\hline
			$K^{+} K^{-}\pi^{-}$                              &$1467    $&$\pm57$&$18.53   $&$\pm0.15$&$15.08 
			$&$\pm0.11$&$14.82   $&$\pm0.11$&$3.34    $&$\pm0.02$&$1.54    $&$\pm0.02$&$1.43    $&$\pm0.01$&$0.92    $&$\pm0.02$\\$K^{+} K^{-}\pi^{-}\pi^{0}$                       &$455     $&$\pm63$&$5.27    $&$\pm0.15$&$5.16  
			$&$\pm0.09$&$4.74    $&$\pm0.09$&$1.16    $&$\pm0.02$&$0.57    $&$\pm0.01$&$0.46    $&$\pm0.01$&$0.23    $&$\pm0.01$\\$\pi^{-}\pi^{+}\pi^{-}$                           &$351     $&$\pm50$&$24.06   $&$\pm0.63$&$18.50 
			$&$\pm0.12$&$21.53   $&$\pm0.13$&$4.06    $&$\pm0.02$&$1.86    $&$\pm0.02$&$2.09    $&$\pm0.01$&$1.45    $&$\pm0.02$\\$K_S^{0} K^{-}$                                   &$310     $&$\pm23$&$22.02   $&$\pm0.35$&$16.73 
			$&$\pm0.14$&$19.21   $&$\pm0.15$&$3.68    $&$\pm0.03$&$1.70    $&$\pm0.02$&$1.81    $&$\pm0.02$&$1.30    $&$\pm0.02$\\$K_S^{0} K^{-}\pi^{0}$                            &$125     $&$\pm32$&$7.54    $&$\pm0.34$&$6.69  
			$&$\pm0.13$&$7.74    $&$\pm0.15$&$1.47    $&$\pm0.03$&$0.69    $&$\pm0.02$&$0.77    $&$\pm0.02$&$0.43    $&$\pm0.02$\\$K^{-}\pi^{-}\pi^{+}$                             &$252     $&$\pm54$&$23.16   $&$\pm1.11$&$16.64 
			$&$\pm0.25$&$18.32   $&$\pm0.27$&$3.64    $&$\pm0.05$&$1.69    $&$\pm0.03$&$1.79    $&$\pm0.03$&$1.21    $&$\pm0.04$\\$K_S^{0} K_S^{0}\pi^{-}$                          &$61      $&$\pm12$&$10.54   $&$\pm0.49$&$7.97  
			$&$\pm0.21$&$9.23    $&$\pm0.23$&$1.80    $&$\pm0.04$&$0.78    $&$\pm0.03$&$0.85    $&$\pm0.02$&$0.52    $&$\pm0.03$\\$K_S^{0} K^{+}\pi^{-}\pi^{-}$                     &$141     $&$\pm19$&$9.94    $&$\pm0.28$&$7.86  
			$&$\pm0.18$&$7.92    $&$\pm0.18$&$1.69    $&$\pm0.03$&$0.78    $&$\pm0.02$&$0.75    $&$\pm0.02$&$0.45    $&$\pm0.02$\\$K_S^{0} K^{-}\pi^{+}\pi^{-}$                     &$119     $&$\pm21$&$8.57    $&$\pm0.43$&$6.78  
			$&$\pm0.13$&$7.16    $&$\pm0.14$&$1.52    $&$\pm0.02$&$0.71    $&$\pm0.01$&$0.68    $&$\pm0.01$&$0.39    $&$\pm0.01$\\$\eta_{\gamma\gamma}\pi^{-}$                      &$160     $&$\pm32$&$17.49   $&$\pm0.77$&$16.08 
			$&$\pm0.11$&$20.53   $&$\pm0.13$&$3.61    $&$\pm0.02$&$1.68    $&$\pm0.02$&$2.01    $&$\pm0.01$&$1.47    $&$\pm0.02$\\$\eta_{\pi^{+}\pi^{-}\pi^{0}}\pi^{-}$             &$26      $&$\pm11$&$9.50    $&$\pm0.55$&$8.73  
			$&$\pm0.09$&$10.54   $&$\pm0.10$&$1.88    $&$\pm0.02$&$0.89    $&$\pm0.01$&$1.02    $&$\pm0.01$&$0.62    $&$\pm0.01$\\$\eta\prime_{\pi^{+}\pi^{-}\eta} \pi^{-}$         &$98      $&$\pm15$&$11.04   $&$\pm0.39$&$8.77  
			$&$\pm0.09$&$10.87   $&$\pm0.10$&$1.94    $&$\pm0.02$&$0.90    $&$\pm0.01$&$1.07    $&$\pm0.01$&$0.68    $&$\pm0.01$\\$\eta\prime_{\gamma\rho^{0}} \pi^{-}$             &$212     $&$\pm34$&$11.29   $&$\pm0.35$&$11.32 
			$&$\pm0.10$&$13.75   $&$\pm0.11$&$2.49    $&$\pm0.02$&$1.21    $&$\pm0.01$&$1.33    $&$\pm0.01$&$0.83    $&$\pm0.01$\\$\eta_{\gamma\gamma}\rho^-$    &$286     $&$\pm73$&$7.65    $&$\pm0.39$&$7.75  
			$&$\pm0.08$&$10.42   $&$\pm0.10$&$1.71    $&$\pm0.02$&$0.81    $&$\pm0.01$&$1.01    $&$\pm0.01$&$0.71    $&$\pm0.01$\\
			\hline\hline
		\end{tabular}
\end{table*}
\begin{table*}[htbp]
	\caption{The  single-tag yields ($N_{\rm ST}$), single-tag efficiencies ($\epsilon_{\rm ST}$), and double-tag efficiencies for each tag mode at 4.290~GeV. The $\epsilon^{\mu_a\nu}_{\rm DT}$ and $\epsilon^{\mu_b\nu}_{\rm DT}$ correspond to the double-tag efficiencies
		for $D^+_s\to \mu^+_a\nu_\mu$ and $D^+_s\to \mu^+_b\nu_\mu$; the $\epsilon^{\tau_e\nu}_{\rm DT}$, $\epsilon^{\tau_\mu\nu}_{\rm DT}$, $\epsilon^{\tau_\pi\nu}_{\rm DT}$, and $\epsilon^{\tau_\rho\nu}_{\rm DT}$ correspond to the double-tag efficiencies for
		$D^{+}_{s} \to \tau^+_e\nu_{\tau}$,
		$D^{+}_{s} \to \tau^+_\mu\nu_{\tau}$,
		$D^{+}_{s} \to \tau^+_\pi\nu_{\tau}$, and
		$D^{+}_{s} \to \tau^+_\rho\nu_{\tau}$, respectively. The efficiencies include the
		BFs of $D_s^{*+}$ and $\tau^+$ decays. The uncertainties are statistical only.
		\label{tab:ST5}}
	\centering
		\begin{tabular}{ l r@{} lr@{}  lr@{} lr@{} lr@{} lr@{} lr@{} lr@{} l }
			\hline\hline
			\makecell{$D^-_s$ tag \\mode}  &\multicolumn{2}{c}{$N_{\rm ST}$} &\multicolumn{2}{c}{\makecell{$\epsilon_{\rm ST}$\\(\%)}}
			&\multicolumn{2}{c}{\makecell{$\epsilon^{\mu_a\nu}_{\rm DT}$\\(\%)}}&\multicolumn{2}{c}{\makecell{$\epsilon^{\mu_b\nu}_{\rm DT}$\\(\%)}}&\multicolumn{2}{c}{\makecell{$\epsilon^{\tau_e\nu}_{\rm DT}$\\(\%)}}&\multicolumn{2}{c}{\makecell{$\epsilon^{\tau_\mu\nu}_{\rm DT}$\\(\%)}}&\multicolumn{2}{c}{\makecell{$\epsilon^{\tau_\pi\nu}_{\rm DT}$\\(\%)}}&\multicolumn{2}{c}{\makecell{$\epsilon^{\tau_\rho\nu}_{\rm DT}$\\(\%)}} \\
			\hline
			$K^{+} K^{-}\pi^{-}$                              &$3432    $&$\pm90$&$18.43   $&$\pm0.09$&$14.34 
			$&$\pm0.08$&$14.86   $&$\pm0.08$&$3.22    $&$\pm0.02$&$1.49    $&$\pm0.01$&$1.40    $&$\pm0.01$&$0.85    $&$\pm0.01$\\$K^{+} K^{-}\pi^{-}\pi^{0}$                       &$1018    $&$\pm81$&$5.02    $&$\pm0.09$&$4.86  
			$&$\pm0.06$&$4.72    $&$\pm0.06$&$1.09    $&$\pm0.01$&$0.53    $&$\pm0.01$&$0.44    $&$\pm0.01$&$0.23    $&$\pm0.01$\\$\pi^{-}\pi^{+}\pi^{-}$                           &$869     $&$\pm87$&$24.33   $&$\pm0.42$&$18.09 
			$&$\pm0.09$&$22.14   $&$\pm0.10$&$4.00    $&$\pm0.02$&$1.85    $&$\pm0.01$&$2.11    $&$\pm0.01$&$1.45    $&$\pm0.01$\\$K_S^{0} K^{-}$                                   &$727     $&$\pm39$&$21.35   $&$\pm0.22$&$15.94 
			$&$\pm0.10$&$19.07   $&$\pm0.11$&$3.51    $&$\pm0.02$&$1.63    $&$\pm0.01$&$1.78    $&$\pm0.01$&$1.24    $&$\pm0.02$\\$K_S^{0} K^{-}\pi^{0}$                            &$239     $&$\pm36$&$7.39    $&$\pm0.22$&$6.28  
			$&$\pm0.09$&$7.68    $&$\pm0.10$&$1.39    $&$\pm0.02$&$0.66    $&$\pm0.01$&$0.73    $&$\pm0.01$&$0.42    $&$\pm0.01$\\$K^{-}\pi^{-}\pi^{+}$                             &$484     $&$\pm76$&$23.61   $&$\pm0.78$&$16.02 
			$&$\pm0.17$&$18.61   $&$\pm0.19$&$3.54    $&$\pm0.03$&$1.63    $&$\pm0.02$&$1.78    $&$\pm0.02$&$1.20    $&$\pm0.03$\\$K_S^{0} K_S^{0}\pi^{-}$                          &$122     $&$\pm19$&$10.12   $&$\pm0.32$&$7.16  
			$&$\pm0.14$&$8.62    $&$\pm0.16$&$1.69    $&$\pm0.03$&$0.80    $&$\pm0.02$&$0.81    $&$\pm0.02$&$0.54    $&$\pm0.02$\\$K_S^{0} K^{+}\pi^{-}\pi^{-}$                     &$370     $&$\pm37$&$8.77    $&$\pm0.17$&$6.77  
			$&$\pm0.12$&$7.48    $&$\pm0.13$&$1.57    $&$\pm0.02$&$0.73    $&$\pm0.02$&$0.71    $&$\pm0.01$&$0.40    $&$\pm0.02$\\$K_S^{0} K^{-}\pi^{+}\pi^{-}$                     &$153     $&$\pm28$&$8.94    $&$\pm0.30$&$6.25  
			$&$\pm0.06$&$6.80    $&$\pm0.07$&$1.42    $&$\pm0.01$&$0.66    $&$\pm0.01$&$0.65    $&$\pm0.01$&$0.37    $&$\pm0.01$\\$\eta_{\gamma\gamma}\pi^{-}$                      &$476     $&$\pm56$&$19.16   $&$\pm0.47$&$15.96 
			$&$\pm0.08$&$20.95   $&$\pm0.09$&$3.52    $&$\pm0.02$&$1.62    $&$\pm0.01$&$2.00    $&$\pm0.01$&$1.45    $&$\pm0.01$\\$\eta_{\pi^{+}\pi^{-}\pi^{0}}\pi^{-}$             &$102     $&$\pm20$&$10.81   $&$\pm0.44$&$8.36  
			$&$\pm0.06$&$10.48   $&$\pm0.07$&$1.89    $&$\pm0.01$&$0.88    $&$\pm0.01$&$0.99    $&$\pm0.01$&$0.63    $&$\pm0.01$\\$\eta\prime_{\pi^{+}\pi^{-}\eta} \pi^{-}$         &$238     $&$\pm25$&$11.13   $&$\pm0.26$&$8.59  
			$&$\pm0.06$&$11.13   $&$\pm0.07$&$1.88    $&$\pm0.01$&$0.89    $&$\pm0.01$&$1.06    $&$\pm0.01$&$0.68    $&$\pm0.01$\\$\eta\prime_{\gamma\rho^{0}} \pi^{-}$             &$462     $&$\pm50$&$13.24   $&$\pm0.34$&$10.89 
			$&$\pm0.07$&$13.99   $&$\pm0.08$&$2.46    $&$\pm0.01$&$1.13    $&$\pm0.01$&$1.33    $&$\pm0.01$&$0.82    $&$\pm0.01$\\$\eta_{\gamma\gamma}\rho^-$    &$624     $&$\pm93$&$7.86    $&$\pm0.27$&$7.39  
			$&$\pm0.06$&$10.30   $&$\pm0.07$&$1.68    $&$\pm0.01$&$0.77    $&$\pm0.01$&$1.01    $&$\pm0.01$&$0.68    $&$\pm0.01$\\
			\hline\hline
		\end{tabular}
\end{table*}
\begin{table*}[htbp]
	\caption{The single-tag yields ($N_{\rm ST}$), single-tag efficiencies ($\epsilon_{\rm ST}$), and double-tag efficiencies for each tag mode at 4.310-4.315~GeV. The $\epsilon^{\mu_a\nu}_{\rm DT}$ and $\epsilon^{\mu_b\nu}_{\rm DT}$ correspond to the double-tag efficiencies
		for $D^+_s\to \mu^+_a\nu_\mu$ and $D^+_s\to \mu^+_b\nu_\mu$; the $\epsilon^{\tau_e\nu}_{\rm DT}$, $\epsilon^{\tau_\mu\nu}_{\rm DT}$, $\epsilon^{\tau_\pi\nu}_{\rm DT}$, and $\epsilon^{\tau_\rho\nu}_{\rm DT}$ correspond to the double-tag efficiencies for
		$D^{+}_{s} \to \tau^+_e\nu_{\tau}$,
		$D^{+}_{s} \to \tau^+_\mu\nu_{\tau}$,
		$D^{+}_{s} \to \tau^+_\pi\nu_{\tau}$, and
		$D^{+}_{s} \to \tau^+_\rho\nu_{\tau}$, respectively. The efficiencies include the
		BFs of $D_s^{*+}$ and $\tau^+$ decays. The uncertainties are statistical only.
		\label{tab:ST6}}
	\centering
		\begin{tabular}{ l r@{} lr@{}  lr@{} lr@{} lr@{} lr@{} lr@{} lr@{} l }
			\hline\hline
			\makecell{$D^-_s$ tag \\mode}  &\multicolumn{2}{c}{$N_{\rm ST}$} &\multicolumn{2}{c}{\makecell{$\epsilon_{\rm ST}$\\(\%)}}
			&\multicolumn{2}{c}{\makecell{$\epsilon^{\mu_a\nu}_{\rm DT}$\\(\%)}}&\multicolumn{2}{c}{\makecell{$\epsilon^{\mu_b\nu}_{\rm DT}$\\(\%)}}&\multicolumn{2}{c}{\makecell{$\epsilon^{\tau_e\nu}_{\rm DT}$\\(\%)}}&\multicolumn{2}{c}{\makecell{$\epsilon^{\tau_\mu\nu}_{\rm DT}$\\(\%)}}&\multicolumn{2}{c}{\makecell{$\epsilon^{\tau_\pi\nu}_{\rm DT}$\\(\%)}}&\multicolumn{2}{c}{\makecell{$\epsilon^{\tau_\rho\nu}_{\rm DT}$\\(\%)}} \\
			\hline
			$K^{+} K^{-}\pi^{-}$                              &$2014    $&$\pm83$&$16.86   $&$\pm0.14$&$12.60 
			$&$\pm0.07$&$13.34   $&$\pm0.08$&$2.80    $&$\pm0.01$&$1.32    $&$\pm0.01$&$1.26    $&$\pm0.01$&$0.81    $&$\pm0.01$\\$K^{+} K^{-}\pi^{-}\pi^{0}$                       &$722     $&$\pm94$&$5.70    $&$\pm0.18$&$4.25  
			$&$\pm0.06$&$4.17    $&$\pm0.06$&$0.97    $&$\pm0.01$&$0.46    $&$\pm0.01$&$0.41    $&$\pm0.01$&$0.21    $&$\pm0.01$\\$\pi^{-}\pi^{+}\pi^{-}$                           &$415     $&$\pm98$&$22.94   $&$\pm0.73$&$16.02 
			$&$\pm0.08$&$19.66   $&$\pm0.09$&$3.52    $&$\pm0.02$&$1.62    $&$\pm0.01$&$1.89    $&$\pm0.01$&$1.36    $&$\pm0.01$\\$K_S^{0} K^{-}$                                   &$443     $&$\pm32$&$19.15   $&$\pm0.28$&$14.22 
			$&$\pm0.09$&$17.12   $&$\pm0.10$&$3.15    $&$\pm0.02$&$1.47    $&$\pm0.01$&$1.60    $&$\pm0.01$&$1.16    $&$\pm0.01$\\$K_S^{0} K^{-}\pi^{0}$                            &$188     $&$\pm46$&$6.63    $&$\pm0.42$&$5.51  
			$&$\pm0.09$&$6.74    $&$\pm0.10$&$1.23    $&$\pm0.02$&$0.59    $&$\pm0.01$&$0.64    $&$\pm0.01$&$0.41    $&$\pm0.01$\\$K^{-}\pi^{-}\pi^{+}$                             &$254     $&$\pm48$&$27.59   $&$\pm1.31$&$14.16 
			$&$\pm0.16$&$16.87   $&$\pm0.18$&$3.15    $&$\pm0.03$&$1.45    $&$\pm0.02$&$1.57    $&$\pm0.02$&$1.12    $&$\pm0.03$\\$K_S^{0} K_S^{0}\pi^{-}$                          &$85      $&$\pm18$&$9.49    $&$\pm0.52$&$6.49  
			$&$\pm0.14$&$7.95    $&$\pm0.15$&$1.51    $&$\pm0.03$&$0.71    $&$\pm0.02$&$0.81    $&$\pm0.02$&$0.54    $&$\pm0.02$\\$K_S^{0} K^{+}\pi^{-}\pi^{-}$                     &$237     $&$\pm42$&$8.31    $&$\pm0.27$&$6.31  
			$&$\pm0.11$&$7.29    $&$\pm0.12$&$1.46    $&$\pm0.02$&$0.70    $&$\pm0.02$&$0.68    $&$\pm0.01$&$0.38    $&$\pm0.02$\\$K_S^{0} K^{-}\pi^{+}\pi^{-}$                     &$156     $&$\pm45$&$9.28    $&$\pm0.58$&$5.86  
			$&$\pm0.06$&$6.47    $&$\pm0.07$&$1.29    $&$\pm0.01$&$0.59    $&$\pm0.01$&$0.60    $&$\pm0.01$&$0.36    $&$\pm0.01$\\$\eta_{\gamma\gamma}\pi^{-}$                      &$211     $&$\pm55$&$15.60   $&$\pm0.71$&$13.77 
			$&$\pm0.08$&$18.24   $&$\pm0.09$&$3.03    $&$\pm0.01$&$1.38    $&$\pm0.01$&$1.74    $&$\pm0.01$&$1.32    $&$\pm0.01$\\$\eta_{\pi^{+}\pi^{-}\pi^{0}}\pi^{-}$             &$70      $&$\pm17$&$8.87    $&$\pm0.62$&$7.33  
			$&$\pm0.06$&$9.40    $&$\pm0.07$&$1.63    $&$\pm0.01$&$0.77    $&$\pm0.01$&$0.87    $&$\pm0.01$&$0.56    $&$\pm0.01$\\$\eta\prime_{\pi^{+}\pi^{-}\eta} \pi^{-}$         &$139     $&$\pm21$&$9.21    $&$\pm0.36$&$7.37  
			$&$\pm0.06$&$9.78    $&$\pm0.07$&$1.66    $&$\pm0.01$&$0.78    $&$\pm0.01$&$0.92    $&$\pm0.01$&$0.64    $&$\pm0.01$\\$\eta\prime_{\gamma\rho^{0}} \pi^{-}$             &$269     $&$\pm55$&$12.80   $&$\pm0.55$&$9.65  
			$&$\pm0.06$&$12.31   $&$\pm0.07$&$2.17    $&$\pm0.01$&$1.01    $&$\pm0.01$&$1.18    $&$\pm0.01$&$0.76    $&$\pm0.01$\\$\eta_{\gamma\gamma}\rho^-$    &$555     $&$\pm102$&$6.23    $&$\pm0.37$&$6.38 
			$&$\pm0.05$&$8.94    $&$\pm0.06$&$1.42    $&$\pm0.01$&$0.67    $&$\pm0.01$&$0.86    $&$\pm0.01$&$0.59    $&$\pm0.01$\\
			\hline\hline
		\end{tabular}
\end{table*}
\begin{table*}[htbp]
	\caption{The  single-tag yields ($N_{\rm ST}$), single-tag efficiencies ($\epsilon_{\rm ST}$), and double-tag efficiencies for each tag mode at 4.400~GeV. The $\epsilon^{\mu_a\nu}_{\rm DT}$ and $\epsilon^{\mu_b\nu}_{\rm DT}$ correspond to the double-tag efficiencies
		for $D^+_s\to \mu^+_a\nu_\mu$ and $D^+_s\to \mu^+_b\nu_\mu$; the $\epsilon^{\tau_e\nu}_{\rm DT}$, $\epsilon^{\tau_\mu\nu}_{\rm DT}$, $\epsilon^{\tau_\pi\nu}_{\rm DT}$, and $\epsilon^{\tau_\rho\nu}_{\rm DT}$ correspond to the double-tag efficiencies for
		$D^{+}_{s} \to \tau^+_e\nu_{\tau}$,
		$D^{+}_{s} \to \tau^+_\mu\nu_{\tau}$,
		$D^{+}_{s} \to \tau^+_\pi\nu_{\tau}$, and
		$D^{+}_{s} \to \tau^+_\rho\nu_{\tau}$, respectively. The efficiencies include the
		BFs of $D_s^{*+}$ and $\tau^+$ decays. The uncertainties are statistical only.
		\label{tab:ST7}}
	\centering
		\begin{tabular}{ l r@{} lr@{}  lr@{} lr@{} lr@{} lr@{} lr@{} lr@{}l }
			\hline\hline
			\makecell{$D^-_s$ tag \\mode}  &\multicolumn{2}{c}{$N_{\rm ST}$} &\multicolumn{2}{c}{\makecell{$\epsilon_{\rm ST}$\\(\%)}}
			&\multicolumn{2}{c}{\makecell{$\epsilon^{\mu_a\nu}_{\rm DT}$\\(\%)}}&\multicolumn{2}{c}{\makecell{$\epsilon^{\mu_b\nu}_{\rm DT}$\\(\%)}}&\multicolumn{2}{c}{\makecell{$\epsilon^{\tau_e\nu}_{\rm DT}$\\(\%)}}&\multicolumn{2}{c}{\makecell{$\epsilon^{\tau_\mu\nu}_{\rm DT}$\\(\%)}}&\multicolumn{2}{c}{\makecell{$\epsilon^{\tau_\pi\nu}_{\rm DT}$\\(\%)}}&\multicolumn{2}{c}{\makecell{$\epsilon^{\tau_\rho\nu}_{\rm DT}$\\(\%)}} \\
			\hline
			$K^{+} K^{-}\pi^{-}$                              &$1544$&$\pm83$&$17.06   $&$\pm0.19$&$12.68   $&$\pm0.07$&$13.80   $&$\pm0.08$&$2.83    $&$\pm0.01$&$1.29    $&$\pm0.0
			1$&$1.26    $&$\pm0.01$&$0.87    $&$\pm0.01$\\
			$K_S^{0} K^{-}$                                   &$311$&$\pm27$&$20.05   $&$\pm0.36$&$14.18   $&$\pm0.09$&$17.15   $&$\pm0.10$&$3.12    $&$\pm0.02$&$1.43    $&$\pm0.01
			$&$1.60    $&$\pm0.01$&$1.24    $&$\pm0.02$\\
			\hline\hline
		\end{tabular}
\end{table*}
\begin{table*}[htbp]
	\caption{The  single-tag yields ($N_{\rm ST}$), single-tag efficiencies ($\epsilon_{\rm ST}$), and double-tag efficiencies for each tag mode at 4.420~GeV. The $\epsilon^{\mu_a\nu}_{\rm DT}$ and $\epsilon^{\mu_b\nu}_{\rm DT}$ correspond to the double-tag efficiencies
		for $D^+_s\to \mu^+_a\nu_\mu$ and $D^+_s\to \mu^+_b\nu_\mu$; the $\epsilon^{\tau_e\nu}_{\rm DT}$, $\epsilon^{\tau_\mu\nu}_{\rm DT}$, $\epsilon^{\tau_\pi\nu}_{\rm DT}$, and $\epsilon^{\tau_\rho\nu}_{\rm DT}$ correspond to the double-tag efficiencies for
		$D^{+}_{s} \to \tau^+_e\nu_{\tau}$,
		$D^{+}_{s} \to \tau^+_\mu\nu_{\tau}$,
		$D^{+}_{s} \to \tau^+_\pi\nu_{\tau}$, and
		$D^{+}_{s} \to \tau^+_\rho\nu_{\tau}$, respectively. The efficiencies include the
		BFs of $D_s^{*+}$ and $\tau^+$ decays. The uncertainties are statistical only.
		\label{tab:ST8}}
	\centering
		\begin{tabular}{ l r@{} lr@{}  lr@{} lr@{} lr@{} lr@{} lr@{} lr@{} l }
			\hline\hline
			\makecell{$D^-_s$ tag \\mode}  &\multicolumn{2}{c}{$N_{\rm ST}$} &\multicolumn{2}{c}{\makecell{$\epsilon_{\rm ST}$\\(\%)}}
			&\multicolumn{2}{c}{\makecell{$\epsilon^{\mu_a\nu}_{\rm DT}$\\(\%)}}&\multicolumn{2}{c}{\makecell{$\epsilon^{\mu_b\nu}_{\rm DT}$\\(\%)}}&\multicolumn{2}{c}{\makecell{$\epsilon^{\tau_e\nu}_{\rm DT}$\\(\%)}}&\multicolumn{2}{c}{\makecell{$\epsilon^{\tau_\mu\nu}_{\rm DT}$\\(\%)}}&\multicolumn{2}{c}{\makecell{$\epsilon^{\tau_\pi\nu}_{\rm DT}$\\(\%)}}&\multicolumn{2}{c}{\makecell{$\epsilon^{\tau_\rho\nu}_{\rm DT}$\\(\%)}} \\
			\hline
			$K^{+} K^{-}\pi^{-}$                              &$5297    $&$\pm138$&$18.44   $&$\pm0.10$&$13.67
			$&$\pm0.08$&$15.07   $&$\pm0.08$&$3.08    $&$\pm0.02$&$1.42    $&$\pm0.01$&$1.39    $&$\pm0.01$&$1.00    $&$\pm0.01$\\$K^{+} K^{-}\pi^{-}\pi^{0}$                       &$1539    $&$\pm218$&$5.36    $&$\pm0.11$&$4.96 
			$&$\pm0.06$&$5.18    $&$\pm0.06$&$1.10    $&$\pm0.01$&$0.51    $&$\pm0.01$&$0.49    $&$\pm0.01$&$0.31    $&$\pm0.01$\\$\pi^{-}\pi^{+}\pi^{-}$                           &$1185    $&$\pm135$&$22.40   $&$\pm0.50$&$17.46
			$&$\pm0.08$&$21.69   $&$\pm0.09$&$3.84    $&$\pm0.02$&$1.76    $&$\pm0.01$&$2.02    $&$\pm0.01$&$1.58    $&$\pm0.01$\\$K_S^{0} K^{-}$                                   &$1195    $&$\pm53$&$21.90   $&$\pm0.19$&$15.85 
			$&$\pm0.10$&$19.25   $&$\pm0.11$&$3.47    $&$\pm0.02$&$1.59    $&$\pm0.01$&$1.80    $&$\pm0.01$&$1.41    $&$\pm0.02$\\$K_S^{0} K^{-}\pi^{0}$                            &$485     $&$\pm84$&$7.81    $&$\pm0.30$&$6.09  
			$&$\pm0.09$&$7.62    $&$\pm0.10$&$1.37    $&$\pm0.02$&$0.64    $&$\pm0.01$&$0.71    $&$\pm0.01$&$0.49    $&$\pm0.01$\\$K^{-}\pi^{-}\pi^{+}$                             &$684     $&$\pm135$&$21.28   $&$\pm0.90$&$15.63
			$&$\pm0.17$&$18.90   $&$\pm0.19$&$3.43    $&$\pm0.03$&$1.55    $&$\pm0.02$&$1.76    $&$\pm0.02$&$1.33    $&$\pm0.03$\\$K_S^{0} K_S^{0}\pi^{-}$                          &$192     $&$\pm35$&$10.32   $&$\pm0.34$&$8.02  
			$&$\pm0.15$&$9.76    $&$\pm0.17$&$1.75    $&$\pm0.03$&$0.84    $&$\pm0.02$&$0.88    $&$\pm0.02$&$0.64    $&$\pm0.02$\\$K_S^{0} K^{+}\pi^{-}\pi^{-}$                     &$588     $&$\pm58$&$9.58    $&$\pm0.19$&$7.51  
			$&$\pm0.12$&$8.56    $&$\pm0.13$&$1.68    $&$\pm0.02$&$0.78    $&$\pm0.02$&$0.80    $&$\pm0.01$&$0.52    $&$\pm0.02$\\$K_S^{0} K^{-}\pi^{+}\pi^{-}$                     &$314     $&$\pm71$&$10.05   $&$\pm0.37$&$6.77  
			$&$\pm0.07$&$7.73    $&$\pm0.07$&$1.50    $&$\pm0.01$&$0.69    $&$\pm0.01$&$0.70    $&$\pm0.01$&$0.46    $&$\pm0.01$\\$\eta_{\gamma\gamma}\pi^{-}$                      &$591     $&$\pm79$&$18.65   $&$\pm0.49$&$14.32 
			$&$\pm0.08$&$19.12   $&$\pm0.09$&$3.13    $&$\pm0.01$&$1.42    $&$\pm0.01$&$1.78    $&$\pm0.01$&$1.44    $&$\pm0.01$\\$\eta_{\pi^{+}\pi^{-}\pi^{0}}\pi^{-}$             &$164     $&$\pm29$&$9.82    $&$\pm0.44$&$7.61  
			$&$\pm0.06$&$9.88    $&$\pm0.07$&$1.69    $&$\pm0.01$&$0.79    $&$\pm0.01$&$0.93    $&$\pm0.01$&$0.66    $&$\pm0.01$\\$\eta\prime_{\pi^{+}\pi^{-}\eta} \pi^{-}$         &$389     $&$\pm30$&$10.13   $&$\pm0.25$&$8.03  
			$&$\pm0.07$&$10.64   $&$\pm0.08$&$1.77    $&$\pm0.01$&$0.82    $&$\pm0.01$&$0.98    $&$\pm0.01$&$0.72    $&$\pm0.01$\\$\eta\prime_{\gamma\rho^{0}} \pi^{-}$             &$949     $&$\pm120$&$13.41   $&$\pm0.36$&$10.44
			$&$\pm0.07$&$13.49   $&$\pm0.08$&$2.32    $&$\pm0.01$&$1.07    $&$\pm0.01$&$1.26    $&$\pm0.01$&$0.89    $&$\pm0.01$\\$\eta_{\gamma\gamma}\rho^-$    &$1318    $&$\pm227$&$7.58    $&$\pm0.26$&$6.32 
			$&$\pm0.05$&$8.87    $&$\pm0.06$&$1.40    $&$\pm0.01$&$0.65    $&$\pm0.01$&$0.84    $&$\pm0.01$&$0.61    $&$\pm0.01$\\
			\hline\hline
		\end{tabular}
\end{table*}
\begin{table*}[htbp]
	\caption{The  single-tag yields ($N_{\rm ST}$), single-tag efficiencies ($\epsilon_{\rm ST}$), and double-tag efficiencies for each tag mode at 4.440~GeV. The $\epsilon^{\mu_a\nu}_{\rm DT}$ and $\epsilon^{\mu_b\nu}_{\rm DT}$ correspond to the double-tag efficiencies
		for $D^+_s\to \mu^+_a\nu_\mu$ and $D^+_s\to \mu^+_b\nu_\mu$; the $\epsilon^{\tau_e\nu}_{\rm DT}$, $\epsilon^{\tau_\mu\nu}_{\rm DT}$, $\epsilon^{\tau_\pi\nu}_{\rm DT}$, and $\epsilon^{\tau_\rho\nu}_{\rm DT}$ correspond to the double-tag efficiencies for
		$D^{+}_{s} \to \tau^+_e\nu_{\tau}$,
		$D^{+}_{s} \to \tau^+_\mu\nu_{\tau}$,
		$D^{+}_{s} \to \tau^+_\pi\nu_{\tau}$, and
		$D^{+}_{s} \to \tau^+_\rho\nu_{\tau}$, respectively. The efficiencies include the
		BFs of $D_s^{*+}$ and $\tau^+$ decays. The uncertainties are statistical only.
		\label{tab:ST9}}
	\centering
		\begin{tabular}{ l r@{} lr@{}  lr@{} lr@{} lr@{} lr@{} lr@{} lr@{} l }
			\hline\hline
			\makecell{$D^-_s$ tag \\mode}  &\multicolumn{2}{c}{$N_{\rm ST}$} &\multicolumn{2}{c}{\makecell{$\epsilon_{\rm ST}$\\(\%)}}
			&\multicolumn{2}{c}{\makecell{$\epsilon^{\mu_a\nu}_{\rm DT}$\\(\%)}}&\multicolumn{2}{c}{\makecell{$\epsilon^{\mu_b\nu}_{\rm DT}$\\(\%)}}&\multicolumn{2}{c}{\makecell{$\epsilon^{\tau_e\nu}_{\rm DT}$\\(\%)}}&\multicolumn{2}{c}{\makecell{$\epsilon^{\tau_\mu\nu}_{\rm DT}$\\(\%)}}&\multicolumn{2}{c}{\makecell{$\epsilon^{\tau_\pi\nu}_{\rm DT}$\\(\%)}}&\multicolumn{2}{c}{\makecell{$\epsilon^{\tau_\rho\nu}_{\rm DT}$\\(\%)}} \\
			\hline
		$K^{+} K^{-}\pi^{-}$                              &$3512    $&$\pm100$&$17.87   $&$\pm0.11$&$13.45
		$&$\pm0.08$&$15.05   $&$\pm0.08$&$3.01    $&$\pm0.01$&$1.37    $&$\pm0.01$&$1.38    $&$\pm0.01$&$0.95    $&$\pm0.01$\\$K^{+} K^{-}\pi^{-}\pi^{0}$                       &$1341    $&$\pm158$&$5.04    $&$\pm0.11$&$4.86 
		$&$\pm0.06$&$5.19    $&$\pm0.06$&$1.09    $&$\pm0.01$&$0.50    $&$\pm0.01$&$0.46    $&$\pm0.01$&$0.28    $&$\pm0.01$\\$\pi^{-}\pi^{+}\pi^{-}$                           &$618     $&$\pm90$&$24.03   $&$\pm0.55$&$17.07 
		$&$\pm0.08$&$21.38   $&$\pm0.09$&$3.82    $&$\pm0.02$&$1.73    $&$\pm0.01$&$2.01    $&$\pm0.01$&$1.56    $&$\pm0.01$\\$K_S^{0} K^{-}$                                   &$769     $&$\pm41$&$21.09   $&$\pm0.23$&$15.50 
		$&$\pm0.09$&$19.01   $&$\pm0.11$&$3.35    $&$\pm0.02$&$1.52    $&$\pm0.01$&$1.73    $&$\pm0.01$&$1.32    $&$\pm0.02$\\$K_S^{0} K^{-}\pi^{0}$                            &$244     $&$\pm67$&$6.79    $&$\pm0.32$&$5.87  
		$&$\pm0.09$&$7.36    $&$\pm0.10$&$1.30    $&$\pm0.02$&$0.60    $&$\pm0.01$&$0.69    $&$\pm0.01$&$0.46    $&$\pm0.01$\\$K^{-}\pi^{-}\pi^{+}$                             &$522     $&$\pm131$&$21.68   $&$\pm1.03$&$15.07
		$&$\pm0.17$&$18.40   $&$\pm0.19$&$3.42    $&$\pm0.03$&$1.56    $&$\pm0.02$&$1.68    $&$\pm0.02$&$1.30    $&$\pm0.03$\\$K_S^{0} K_S^{0}\pi^{-}$                          &$156     $&$\pm26$&$9.64    $&$\pm0.38$&$7.34  
		$&$\pm0.14$&$9.08    $&$\pm0.16$&$1.66    $&$\pm0.03$&$0.72    $&$\pm0.02$&$0.83    $&$\pm0.02$&$0.55    $&$\pm0.02$\\$K_S^{0} K^{+}\pi^{-}\pi^{-}$                     &$362     $&$\pm38$&$9.15    $&$\pm0.20$&$7.08  
		$&$\pm0.12$&$8.00    $&$\pm0.13$&$1.54    $&$\pm0.02$&$0.72    $&$\pm0.02$&$0.73    $&$\pm0.01$&$0.49    $&$\pm0.02$\\$K_S^{0} K^{-}\pi^{+}\pi^{-}$                     &$216     $&$\pm37$&$8.22    $&$\pm0.38$&$6.28  
		$&$\pm0.06$&$7.20    $&$\pm0.07$&$1.43    $&$\pm0.01$&$0.64    $&$\pm0.01$&$0.66    $&$\pm0.01$&$0.42    $&$\pm0.01$\\$\eta_{\gamma\gamma}\pi^{-}$                      &$332     $&$\pm76$&$16.88   $&$\pm0.85$&$13.94 
		$&$\pm0.08$&$18.81   $&$\pm0.09$&$3.10    $&$\pm0.01$&$1.39    $&$\pm0.01$&$1.75    $&$\pm0.01$&$1.43    $&$\pm0.02$\\$\eta_{\pi^{+}\pi^{-}\pi^{0}}\pi^{-}$             &$134     $&$\pm23$&$9.52    $&$\pm0.50$&$7.49  
		$&$\pm0.06$&$9.84    $&$\pm0.07$&$1.65    $&$\pm0.01$&$0.75    $&$\pm0.01$&$0.90    $&$\pm0.01$&$0.64    $&$\pm0.01$\\$\eta\prime_{\pi^{+}\pi^{-}\eta} \pi^{-}$         &$213     $&$\pm21$&$9.83    $&$\pm0.30$&$7.87  
		$&$\pm0.06$&$10.50   $&$\pm0.07$&$1.73    $&$\pm0.01$&$0.77    $&$\pm0.01$&$0.96    $&$\pm0.01$&$0.72    $&$\pm0.01$\\$\eta\prime_{\gamma\rho^{0}} \pi^{-}$             &$608     $&$\pm68$&$12.63   $&$\pm0.44$&$10.25 
		$&$\pm0.07$&$13.36   $&$\pm0.08$&$2.28    $&$\pm0.01$&$1.05    $&$\pm0.01$&$1.24    $&$\pm0.01$&$0.87    $&$\pm0.01$\\$\eta_{\gamma\gamma}\rho^-$    &$672     $&$\pm129$&$6.09    $&$\pm0.36$&$6.16 
		$&$\pm0.05$&$8.70    $&$\pm0.06$&$1.38    $&$\pm0.01$&$0.63    $&$\pm0.01$&$0.82    $&$\pm0.01$&$0.61    $&$\pm0.01$\\
			\hline\hline
		\end{tabular}
\end{table*}
\begin{table*}[htbp]
	\caption{The  single-tag yields ($N_{\rm ST}$), single-tag efficiencies ($\epsilon_{\rm ST}$), and double-tag efficiencies for each tag mode at 4.470-4.699~GeV. The $\epsilon^{\mu_a\nu}_{\rm DT}$ and $\epsilon^{\mu_b\nu}_{\rm DT}$ correspond to the double-tag efficiencies
		for $D^+_s\to \mu^+_a\nu_\mu$ and $D^+_s\to \mu^+_b\nu_\mu$; the $\epsilon^{\tau_e\nu}_{\rm DT}$, $\epsilon^{\tau_\mu\nu}_{\rm DT}$, $\epsilon^{\tau_\pi\nu}_{\rm DT}$, and $\epsilon^{\tau_\rho\nu}_{\rm DT}$ correspond to the double-tag efficiencies for
		$D^{+}_{s} \to \tau^+_e\nu_{\tau}$,
		$D^{+}_{s} \to \tau^+_\mu\nu_{\tau}$,
		$D^{+}_{s} \to \tau^+_\pi\nu_{\tau}$, and
		$D^{+}_{s} \to \tau^+_\rho\nu_{\tau}$, respectively. The efficiencies include the
		BFs of $D_s^{*+}$ and $\tau^+$ decays. The uncertainties are statistical only.
		\label{tab:ST10}}
	\centering
		\begin{tabular}{ l r@{} lr@{}  lr@{} lr@{} lr@{} lr@{} lr@{}  lr@{}l }
			\hline\hline
			\makecell{$D^-_s$ tag \\mode}  &\multicolumn{2}{c}{$N_{\rm ST}$} &\multicolumn{2}{c}{\makecell{$\epsilon_{\rm ST}$\\(\%)}}
			&\multicolumn{2}{c}{\makecell{$\epsilon^{\mu_a\nu}_{\rm DT}$\\(\%)}}&\multicolumn{2}{c}{\makecell{$\epsilon^{\mu_b\nu}_{\rm DT}$\\(\%)}}&\multicolumn{2}{c}{\makecell{$\epsilon^{\tau_e\nu}_{\rm DT}$\\(\%)}}&\multicolumn{2}{c}{\makecell{$\epsilon^{\tau_\mu\nu}_{\rm DT}$\\(\%)}}&\multicolumn{2}{c}{\makecell{$\epsilon^{\tau_\pi\nu}_{\rm DT}$\\(\%)}}&\multicolumn{2}{c}{\makecell{$\epsilon^{\tau_\rho\nu}_{\rm DT}$\\(\%)}} \\
			\hline
			$K^{+} K^{-}\pi^{-}$                              &$10840$&$\pm266$&$15.44   $&$\pm0.07$&$11.12   $&$\pm0.03$&$12.85   $&$\pm0.03$&$2.57    $&$\pm0.01$&$1.14    $&$\pm0
			.01$&$1.12    $&$\pm0.01$&$0.84    $&$\pm0.01$\\$K^{+} K^{-}\pi^{-}\pi^{0}$                       &$4412$&$\pm487$&$4.99    $&$\pm0.11$&$4.24    $&$\pm0.02$&$4.77    $&$\pm0.03$&$0.98    $&$\pm0.01$&$0.44    $&$\pm0.   01$&$0.41    $&$\pm0.01$&$0.26    $&$\pm0.01$\\$\pi^{-}\pi^{+}\pi^{-}$                            &$2870$&$\pm383$&$19.22   $&$\pm0.45$&$13.73   $&$\pm0.03$&$17.63   $&$\pm0.04$&$3.16    $&$\pm0.01$&$1.38    $&$\pm0.   01$&$1.56    $&$\pm0.01$&$1.25    $&$\pm0.01$\\$K_S^{0} K^{-}$                                    &$2217$&$\pm91$&$17.41   $&$\pm0.15$&$12.33   $&$\pm0.04$&$15.44   $&$\pm0.04$&$2.82    $&$\pm0.01$&$1.22    $&$\pm0.0  1$&$1.35    $&$\pm0.01$&$1.10    $&$\pm0.01$\\$K_S^{0} K^{-}\pi^{0}$                              &$643$&$\pm158$&$5.56    $&$\pm0.24$&$4.91    $&$\pm0.04$&$6.22    $&$\pm0.04$&$1.12    $&$\pm0.01$&$0.50    $&$\pm0.0  1$&$0.55    $&$\pm0.01$&$0.40    $&$\pm0.01$\\$K_S^{0} K_S^{0}\pi^{-}$                            &$346$&$\pm68$&$7.77    $&$\pm0.26$&$6.08    $&$\pm0.06$&$7.72    $&$\pm0.07$&$1.39    $&$\pm0.01$&$0.63    $&$\pm0.01 $&$0.68    $&$\pm0.01$&$0.47    $&$\pm0.01$\\$K_S^{0} K^{+}\pi^{-}\pi^{-}$                        &$1114$&$\pm97$&$8.01    $&$\pm0.14$&$6.03    $&$\pm0.05$&$7.07    $&$\pm0.05$&$1.38    $&$\pm0.01$&$0.60    $&$\pm0.0 1$&$0.60    $&$\pm0.01$&$0.42    $&$\pm0.01$\\$K_S^{0} K^{-}\pi^{+}\pi^{-}$                       &$376$&$\pm90$&$8.10    $&$\pm0.30$&$5.38    $&$\pm0.03$&$6.33    $&$\pm0.03$&$1.24    $&$\pm0.01$&$0.55    $&$\pm0.01 $&$0.54    $&$\pm0.01$&$0.37    $&$\pm0.01$\\$\eta\prime_{\pi^{+}\pi^{-}\eta} \pi^{-}$            &$735$&$\pm87$&$8.17    $&$\pm0.22$&$6.30    $&$\pm0.02$&$8.62    $&$\pm0.03$&$1.44    $&$\pm0.00$&$0.64    $&$\pm0.01 $&$0.76    $&$\pm0.01$&$0.60    $&$\pm0.01$\\$\eta\prime_{\gamma\rho^{0}} \pi^{-}$                &$1603$&$\pm252$&$10.26   $&$\pm0.30$&$8.16    $&$\pm0.03$&$10.91   $&$\pm0.03$&$1.90    $&$\pm0.01$&$0.83    $&$\pm0.   01$&$0.96    $&$\pm0.01$&$0.72    $&$\pm0.01$\\
			\hline\hline
		\end{tabular}
\end{table*}

\end{document}

%% file: authorlist_2024-01-03.tex
\author{
		M.~Ablikim$^{1}$, M.~N.~Achasov$^{4,c}$, P.~Adlarson$^{75}$, O.~Afedulidis$^{3}$, X.~C.~Ai$^{80}$, R.~Aliberti$^{35}$, A.~Amoroso$^{74A,74C}$, Q.~An$^{71,58,a}$, Y.~Bai$^{57}$, O.~Bakina$^{36}$
% ...
		\\
		\vspace{0.2cm}
		(BESIII Collaboration)\\
		\vspace{0.2cm} {\it
			$^{1}$ Institute of High Energy Physics, Beijing 100049, People's Republic of China\\
% ...
$^{3}$ Bochum Ruhr-University, D-44780 Bochum, Germany\\
			$^{4}$ Budker Institute of Nuclear Physics SB RAS (BINP), Novosibirsk 630090, Russia\\
% ...
$^{35}$ Johannes Gutenberg University of Mainz, Johann-Joachim-Becher-Weg 45, D-55099 Mainz, Germany\\
			$^{36}$ Joint Institute for Nuclear Research, 141980 Dubna, Moscow region, Russia\\
% ...
$^{57}$ Southeast University, Nanjing 211100, People's Republic of China\\
			$^{58}$ State Key Laboratory of Particle Detection and Electronics, Beijing 100049, Hefei 230026, People's Republic of China\\
% ...
$^{71}$ University of Science and Technology of China, Hefei 230026, People's Republic of China\\
% ...
$^{74}$ University of Turin and INFN, (A)University of Turin, I-10125, Turin, Italy; (B)University of Eastern Piedmont, I-15121, Alessandria, Italy; (C)INFN, I-10125, Turin, Italy\\
			$^{75}$ Uppsala University, Box 516, SE-75120 Uppsala, Sweden\\
% ...
$^{80}$ Zhengzhou University, Zhengzhou 450001, People's Republic of China\\
			\vspace{0.2cm}
			$^{a}$ Deceased\\
% ...
$^{c}$ Also at the Novosibirsk State University, Novosibirsk, 630090, Russia\\
% ...
	}
}